\documentclass[journal]{IEEEtran}

\usepackage{setspace}
\usepackage{lettrine}

\usepackage{graphics,
           psfrag,
           epsfig,
           amsthm,
           cite,
           amssymb,
           url,
           dsfont,
           subfigure,
           balance,
           enumerate,
           color,
           setspace
}
\usepackage{amsmath}

\usepackage{bm}
\usepackage{epstopdf}
\usepackage{algorithm} 
\usepackage{algorithmicx}
\usepackage{algpseudocode}

\newcommand{\eref}[1]{(\ref{#1})}
\newcommand{\sref}[1]{Section~\ref{#1}}

\newcommand{\cref}[1]{Constraint~\ref{#1}}

\newcommand{\ignore}[1]{}

\IEEEoverridecommandlockouts
\usepackage{mathtools}
\usepackage{color}
\usepackage[utf8]{inputenc}

\begin{document}
\IEEEoverridecommandlockouts


\title{\vspace{-.9cm} A Joint Reinforcement-Learning Enabled Caching and Cross-Layer Network Code for Sum-Rate Maximization in F-RAN with D2D Communications}

\author{
 \IEEEauthorblockN{Mohammed S. Al-Abiad, \textit{Student Member, IEEE}, Md. Zoheb Hassan, \textit{Student Member, IEEE}, and Md. Jahangir Hossain, \textit{Senior Member, IEEE}}

\thanks {
Mohammed S. Al-Abiad and Md. Jahangir Hossain are with the School of
Engineering, University of British Columbia, Kelowna, BC V1V 1V7, Canada
(e-mail: m.saif@alumni.ubc.ca, jahangir.hossain@ubc.ca).

Md. Zoheb Hassan is with $\acute{\text{E}}$cole de technologie sup$\acute{\text{e}}$rieure (ETS), University of Quebec, Canada (e-mail:  md-zoheb.hassan.1@ens.etsmtl.ca).
}

}

\maketitle

\begin{abstract}

In this paper, we leverage reinforcement learning (RL) and cross-layer network coding (CLNC) for efficiently pre-fetching users' contents to the local caches and delivering these contents to users in a downlink fog-radio access network (F-RAN) with device-to-device (D2D) communications. In the considered system, fog access points (F-APs) and cache-enabled D2D (CE-D2D) users are equipped with local caches for alleviating traffic burden at the fronthaul, while users' contents can be easily and quickly accommodated. In CLNC, the coding decisions take users' contents, their rates, and power levels of F-APs and CE-D2D users into account, and RL optimizes caching strategy. Towards this
goal, a joint content placement and delivery problem is formulated as an optimization problem with a goal to maximize system sum-rate. For this NP-hard problem, we first develop an innovative decentralized CLNC coalition formation (CLNC-CF) algorithm to obtain a stable solution for the content delivery problem, where F-APs and CE-D2D users  utilize CLNC resource allocation. By taking the behavior
of F-APs and CE-D2D users into account, we then develop a multi-agent RL (MARL) algorithm for optimizing the content placements at both F-APs and  CE-D2D users. Simulation results show that the proposed joint CLNC-CF and RL framework can effectively improve the sum-rate by up to 30\%, 60\%, and 150\%, respectively, compared to: 1) an optimal	uncoded algorithm, 2) a standard rate-aware-NC algorithm, and 3) a benchmark classical NC with network-layer optimization.

\end{abstract}

\begin{IEEEkeywords}

Caching, D2D communications, F-RAN, NC, resource and power allocation, reinforcement learning.

\end{IEEEkeywords}

\section{Introduction} \label{sec:I}
\vspace{-0.1cm}
\IEEEPARstart{F}{og} radio access network (F-RAN) has recently given
significant attention for beyond-5G era while leveraging
the centralized processing of a cloud-RAN (C-RAN) and intelligence of the network edge. In addition, F-RAN takes advantage of a fast access of the contents through distributed local caches at the fog access points (F-APs) \cite{4,5}. Prior F-RAN systems, the particular popular contents were streamed from cloud base stations (CBSs) to network edge \cite{2n}. Essentially, these popular contents require duplicated downloads from the CBS, and such duplicate downloads severely degrade
system performance. By caching the popular contents in the F-APs, the demands from users can be accommodated easily with minimum intervention of  CBSs. Hence, F-RAN significantly alleviates traffic burden at the fronthaul and improves system performance \cite{6}. However, in beyond 5G networks, caching the increased popular contents at F-APs, owing to their  equipment cost and size issue, is a key concern. 

In order to overcome this concern, distributed caching is envisioned, and caching at cache-enabled device-to-device (CE-D2D) users is employed \cite{6NN, 6NNN}. Therefore, the performance of an F-RAN can be further improved by implementing D2D communications \cite{8}, where caching at F-APs and CE-D2D users are leveraged \cite{9}. With the distributed caching in an F-RAN, F-APs and CE-D2D users transmit their cached contents to the interested users via cellular and D2D links, respectively. 
However,  pre-fetching popular contents to F-APs and CE-D2D users for effective delivery needs a careful optimization. Evidently, a significant portion of these popular contents are delay-sensitive, and it is crucial to efficiently pre-fetch them to F-APs and CE-D2D users for immediate and effective delivery. Therefore, developing  an innovative content placement and content delivery framework is imperative for harnessing
the aforementioned benefits of F-RANs.

The joint content placement and content delivery optimization problem in F-RAN has been investigated separately. Particularly, the content delivery-based network coding (NC) \cite{9n} problem was solved using Random Linear NC (RLNC) \cite{10, 11, 12} and Instantly Decodable NC (IDNC) \cite{13,14,15,16,17,18,21,22,25,29,30,31, 32}. RLNC offers an optimal throughput maximization \cite{10}, but it is not a suitable 
technique for delivering delay-sensitive contents for real-time 
applications that require instant content decoding. In contrast, 
IDNC \cite{13,14,15,16,17,18,21,22,25,29,30,31, 32} offers an immediate content delivery, and consequently,
it provides fast and instantaneous decoding process that is affordable for real-time applications, e.g., streaming applications. In the contemporary literature, the works in \cite{33,34,35,36,37} developed simple caching schemes for content placement optimization problem in small network settings.  However, these works did not harvest
the  benefits of IDNC to multiplex many users to the same resource block. Unlike all the aforementioned works, our work considers a joint optimization of content placement and content delivery problem that will be referred as \textit{sum-rate maximization} problem. \ignore{To this end, we consider a downlink F-RAN system with D2D communications comprising
several single-antenna F-APs connected to the CBS using capacity-limited fronthaul links and several CE-D2D users. The F-APs and CE-D2D users that are equipped with local caches implement RL algorithm to optimally cache users' requests and thus, deliver them to users via cellular and D2D links, respectively, with the maximum sum-rate.}   

\ignore{Our work considers the downlink of D2D-aided F-RANs comprising
several single-antenna FAPs connected to the cloud base station (CBS) and several CE-D2D users. The FAPs and CE-D2D users are employed by local caches that implement RL algorithm to optimally cache users requests and thus, deliver them to users via cellular links and D2D links, respectively, with the maximum sum-rate.   The sum-rate
optimization problem is motivated by real-time applications,
i.e., video streaming. In such applications, users need to
immediately stream . Therefore, the main objective
here is to immediately stream a set of popular files to users
in an F-RAN resource setting so as the CBS offloading is
improved while guaranteeing the QoS of users.}

\subsection{Related Works and Motivation}
Most existing and relevant works on
F-RANs focused on user scheduling problem in order to
maximize sum-rate, e.g., \cite{38,39,40}. Specifically, the study in \cite{38} included power allocation optimization
for the F-APs to further improve the sum-rate. However, all of these works viewed the network solely from the physical-layer perspective without taking into consideration upper-layer facts, e.g., combining users’ requests. As a result, only a single user was assigned to each F-AP, and it is not affordable for large scale network. It has been noticed that users tend to stream a popular video, and consequently, users have a common interest in requesting same content within a small interval of time. This frequently  happens in a hotspot, e.g., a playground, a public transport, a conference hall. Actually, transmiting requested contents to users without being combined severely degrades system performance. Therefore, IDNC \cite{13} can wisely select a combination of contents (i.e., using the binary XoR combination) that can multiplex a subset of interested users to the same resource block.

In IDNC-based networks, the content delivery problem was
investigated for various wireless networking scenarios, e.g., point-to-multipoint (PMP) \cite{13,14}, D2D networks \cite{15,16,17},
D2D F-RANs \cite{18}. For example, in \cite{18}, the authors developed a centralized D2D F-RAN scheme for completion time reduction. Unfortunately, the above related works primarily relied on optimization at the network
layer, and their main limitation is that the transmission rate of each F-AP is selected according to the user with
the weakest channel quality. This is inefficient becasue the minimum selected transmission rate results in the prolonged file
reception time and thus, consumes network time resources. To this end, two advancements aimed at developing promising techniques for improving content delivery, namely, (i) rate aware-IDNC (RA-IDNC) and (ii) cross-layer NC (CLNC) schemes.

\textit{Related works used RA-IDNC scheme:} In RA-IDNC, the coding decisions depend on content combinations at the network layer and transmission rate at the physical layer. Such scheme was first proposed in \cite{21} for completion time minimization in PMP system.  In \cite{25}, the authors used RA-IDNC in a practical and promising paradigm of C-RANs. 
However, the authors assumed that all F-APs use a fixed transmit power level. Moreover, for synchronization purpose, the same transmission rate (i.e., the lowest transmission rate of all F-APs as in \cite{25}) is selected, and it is impractical. In fact, it violates the QoS rate guarantee and leads to a longer time for content delivery. In addition, the aforementioned RA-IDNC works ignored the potentials of D2D communications. Addressing these RA-IDNC limitations is imperative for harvesting
the benefits of next-generation F-RANs.

\textit{Related Works used CLNC Scheme:} The authors in \cite{29}, developed CLNC scheme to optimize the employed rates in RA-IDNC decisions using power control on each F-AP in C-RANs. Essentially, the coding decisions in CLNC scheme not only depends on NC and users' rate, but also on the power levels of each F-AP. Accordingly, CLNC is a promising technique for significantly improving sum-rate \cite{30}, cloud offloading \cite{31}, and delay  \cite{32}.  Particularly, the authors developed CLNC schemes for cloud offloading maximization in F-RAN \cite{31} and for delay minimization in D2D-aided F-RAN \cite{32}. However, all of these CLNC works ignored the content placement optimization problem that draws its importance from improving the delivery time of contents to users. More importantly, these works ignored the decision-making
capability of F-APs and CE-D2D users, and proposed  a graph-theoretical centralized
solution. Note that in CLNC-based dense network, the CBS requires to generate  a vertex for each possible NC combination, and thus, its complexity is exponentially increased \cite{29}. Essentially, for a large scale network, these centralized CLNC solutions may not be affordable in practice.

The emerging distributed machine-learning (ML) based algorithms
are postulated to address issues of the conventional
optimization-based resource allocation algorithms. Specifically,
by training a neural-network offline with large samples, ML-based
algorithms allow the radio resource controller to rapidly determine resource allocation decisions while requiring low signaling overhead
\cite{ML_1}. Recently, ML-based resource
optimizations have been developed for caching and power allocation in F-RAN \cite{ML_3,ML_4}.
In our considered optimization problem, the caching decision depends on the resource allocation at the physical-layer, and hence, without knowing that we can not determine caching optimally. Unfortunately, finding the resource allocation decision also depends on caching problem. Therefore, the overall optimization is very complex and can only be solved optimally via exahaustive search. However, this is not feasible. Therefore, distributed RL is a suitable platform becasue it does not need any prior information about the system (i.e., the resource allocation decision).  To the
best of our knowledge, a joint distributed framework of CLNC resource
allocation, D2D communications, and RL caching optimization for F-RAN architecture has not been considered in state-of-the-art literature.\ignore{ To this end, this work  introduces a novel framework that jointly exploits distributed and CLNC resource allocation to maximize sum-rate for F-RAN system.} 

\textit{Motivation:}\ignore{Few years ago users would tolerate some delays for a
video to be streamed before watching. This is almost unacceptable for any
person nowadays with the advancements of wireless communications and
diversity of mobile multimedia applications. Therefore, the design of wireless communication systems of beyond-5G era are efficiently considered high-rate demand, maximum information flow, and immediate content decoding to meet the delay requirements and ensure good streaming quality.} The sum-rate
maximization optimization problem is motivated by real-time applications,
i.e., video streaming, where users need to
immediately stream their requests while ensuring the minimum required quality-of-service (QoS). For this, CLNC can stream users' requests from F-APs and CE-D2D users  with the maximum possible sum-rate, while ensuring fast access to these requests. Essentially, streaming users' requests can be done either by pre-loading them at F-APs and CE-D2D users  at much lower rates or at off-peak times. Unlike these impractical scenarios, our work considers that users’ combined requests
can be delivered with the maximum possible sum-rate, while users can progressively and instantly
use them.

\subsection{ Contributions}
The main contribution of this work is, thus, an innovative CLNC and RL framework jointly taking caching strategy, NC, D2D communications, power/rate optimization, and fronthaul capacity into account. Our key contributions include.
\begin{itemize}
		\item In F-RAN with D2D communications, the F-APs and CE-D2D users exploit their cached files to use CLNC to maximize
	   sum-rate. Hence, our framework jointly considers cache resources optimization and wireless edge communications for content delivery.  To this end, a sum-rate maximization optimization problem is formulated with the constraints on content placement at local caches, NC, users scheduling, their limited coverage zones, transmission rate/power, and QoS rate guarantee. Such optimization problem is NP-hard
	   and computationally intractable.
	   
	\item To solve this joint
	optimization problem, two-stage iterative approach is developed by decomposing user clustering
	and cache resource allocations using multi-leader-follower Stackelberg game. Specifically, we first develop an innovative decentralized CLNC coalition formation (CLNC-CF) algorithm to obtain a stable F-AP and CE-D2D user coalition formation for the followers, where F-APs and CE-D2D users are utilizing CLNC resource allocation. By taking the behavior of F-APs and CE-D2D users into account, we then develop a multi-agent RL (MARL) algorithm for the leaders.
    The aforementioned two-stage of CLNC-CF and MARL solutions is referred to a joint CLNC-CF-RL approach.
	\item We rigorously analyze the convergence and the computational complexity of our joint proposed approach. Specifically, our proposed CLNC-CF algorithm is
	proved to be a Nash-stable and our proposed MARL algorithm is the best strategy for content placement.
	\item Extensive	simulations are conducted to verify performance	gain of our proposed CLNC-CF-RL framework over several
	benchmark schemes. Simulation results show that our proposed joint  CLNC-CF-RL framework can effectively improve the sum-rate by around 30\%, 60\%, and 150\%, respectively, compared to: 1) an optimal	uncoded algorithm, 2) a standard RA-IDNC algorithm, and 3) a benchmark classical IDNC with network-layer optimization. 
\end{itemize}

The rest of this paper is organized as follows. The system model and the CLNC are described  in  \sref{SMMM}. We formulate the sum-rate maximization problem in \sref{PF}. In \sref{JA}, we develop a joint CLNC-CF-RL approach for solving the problem. In \sref{PP}, we analyze
the properties of the developed joint approach. Simulation results are presented in
\sref{NR}, and in \sref{C}, we conclude the paper.

\section{System Overview and CLNC} \label{SMMM}

\subsection{System Model}
\begin{figure}[t!]
	\centering
	\includegraphics[width=0.75\linewidth]{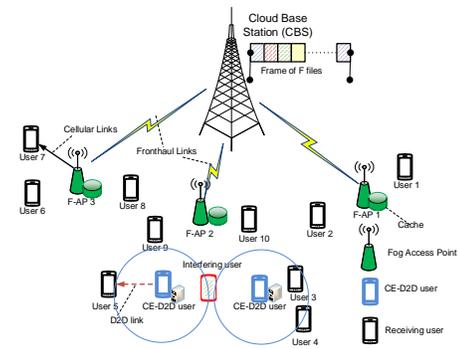}
	\caption{F-RAN model with D2D communications with $11$ users, $3$ F-APs, $3$ CE-D2D users, and $1$ CBS.}
	\label{fig1}
\end{figure}

We consider a downlink F-RAN system with D2D communications illustrated in Fig. \ref{fig1} with one CBS, $K$ F-APs, and $N$ CE-D2D users. The sets of F-APs and CE-D2D users are denoted
by  $\mathcal{K}=\{1,2,\cdots,K\}$ and $\mathcal{N}=\{1,2,\cdots,N\}$, respectively, and they cooperate with each other to serve single-antenna $U$ users, denoted by the set $\mathcal{U}=\{1,2, \cdots,U\}$. Each F-AP is connected to the CBS by a fronthaul link of capacity $C_{fh}$ bits per second. We assumed that each user is equipped with a single antenna and uses half-duplex channel, and accordingly, each user can be served from CE-D2D user via D2D link or from F-AP using cellular channel. Moreover, the allocated channels for D2D communications  are
out-of-band to those used by F-APs, i.e., an overlay D2D communications
model is adopted \cite{9}. We adopt a partially connected D2D networks where
CE-D2D users are low-complexity devices and they can transmit to users at a certain amount of power. Accordingly, each CE-D2D user has limited transmission range, denoted by $\mathcal A_{n}$, which represents the service area of the $n$-th CE-D2D user to transmit data within a circle of radius $\mathtt R$. The set of users  within the transmission range of the $n$-th CE-D2D is defined by $\mathcal A_{n}=\{u\in \mathcal{U}| d_{n,u}\leq \mathtt R$\}, where $d_{n,u}$ is the distance between the $n$-th CE-D2D user and  the $u$-th user. Similar to \cite{32}, we consider that CE-D2D users adopt the same frequency band and serve users simultaneously via D2D links.

Let $\mathcal F$ denote a frame of $F$ files, $\mathcal{F} =\{1,2,\ \cdots, F\}$, each of size $L$ bits. The frame represents a popular content due to its previous multiple downloads by different subsets of users over a short period of time. The frame is
fixed for the considered time period, and its entirely available at the CBS, whereas the F-APs and CE-D2D users can cache up to $\mu F$, where $0\leq \mu \leq 1$ is the fractional cache size. Specifically, each F-AP $k$ and each CE-D2D user $n$ are equipped with local caches, which can proactively cache a subset of files $\mathcal{F}_k$ and $\mathcal{F}_n$ that represent the $k$-th F-AP and the $n$-th CE-D2D user \textit{cache} sets, respectively.

The file placement phase starts by proactively caching popular files to F-APs and CE-D2D users, subject to the aforementioned cache capacity constraints. After the placement phase, the system enters the file delivery phase, which is done in F-AP and CE-D2D user transmissions. We assume that each user is already downloaded a set of the $F$ files, which is denoted by the \textit{Has} set $\mathcal H_u$ of the $u$-th user. At any transmission, each user is arbitrarily interested in streaming one of the $F$ files. The request of the $u$-th user in a given transmission is denoted by the \textit{Wants} set $\mathcal W_u$. For a given file delivery phase, F-APs and CE-D2D users exploit the users' downloaded files to perform XoR encoding operation, respectively, when new files requested by users. At the users, users exploit the downloaded files to extract the requested files immediately by performing XoR decoding operation.\ignore{  Our goal is to optimally place the requested files at the local caches of F-APs and CE-D2D users to be efficiently delivered to the users with a maximum sum-rate by leveraging NC, F-RAN system, and D2D communications.} The notations used throughout
this paper are listed in Table \ref{table_1}.

\begin{table*}[t!]
	\renewcommand{\arraystretch}{0.9}
	\caption{Main Symbols used in the paper}
	\label{table_1}
	\centering
	\begin{tabular}{|p{2.1cm}| p{10.8cm}|}
		\hline
		\textbf{Notation} & \textbf{Description}\\
		\hline
		\hline
		$\mathcal{U}, \mathcal N, \mathcal K, \mathcal F$ & Sets of $U$ users, $N$ CE-D2D users, $K$ F-APs, $F$ popular files \\
		\hline
		$\mathcal U_m$ & Set of users whose requests being missed at the network\\
		\hline
		$\mathcal K_m$ & Set of F-APs serve users in  $\mathcal U_m$\\
		\hline
	  $\mathcal{R}$ & Set of all achievable capacities\\
		\hline
		$\mathcal A_n$ & Set of users in the transmission range of CE-D2D user $n$\\
		\hline
		$\mathcal{W}_{u}(\mathcal H_u)$ & Set of wanted (received) files by user $u$\\
		\hline
		$C_{fh}$ & Capacity of fronthaul link in bits per second \\
		\hline
		$R_{k}$, $C_k$ & Transmission rate of F-AP $k$ and CE-D2D user $n$\\
		\hline
		$\kappa_k, \mathtt f_n$ & The encoded file of F-AP $k$ and CE-D2D user $n$\\
		\hline
		$\tau (\kappa_k), \mathtt u(\mathtt f_n)$ & Set of targeted users by F-AP $k$ and CE-D2D user $n$\\
		\hline
		$\mathtt R_k$ & Data rate of F-AP $k$ for fetching files from the CBS\\
		\hline
	$\mathbf C, \mathbf S, \mathbf H$ & Caching, side information, and channel gain matrices\\
		    \hline
	    $\oplus, \mathcal O$ & XOR operation and complexity notation\\
		\hline
		$x_{u,f}$ & Binary variable representing user $u$ requests file $f$ \\
		\hline 
		$y_{f,k}, y_{f,n}$ & Binary variables representing file $f$ is stored at the caches of F-AP $k$ and CE-D2D user $n$\\
		\hline
		$\Phi, Z$ & F-AP cluster and CE-D2D cluster\\
		\hline
		$\Psi, \mathbf Z$ & Sets of $\Phi$ and $Z$ clusters\\
		\hline
		$\mathcal G_F, >_u, \mathcal G$ & Game, preference, and graph notations\\
		\hline
		$\Gamma$ & History profile\\
		\hline
		$\lambda_t, \alpha_t$ & Learning rates at iteration index $t$\\
		\hline
		 $a_{i,j}$ & Action state of agent $i$ for decision $j$\\
		 \hline
		$x^t_{i,j}$, $\rho^t_{i,j}$&  The estimated
		utility and probability of taking action $a_{i,j}$\\
		\hline
	\end{tabular}
\end{table*}

\subsection{Cross-layer Modeling: CLNC} \label{CLNC}
\ignore{ In this subsection, we describe the cross-layer architecture (i.e., physical and network layers) of the D2D-aided F-RAN system.first, followed by  caching and file delivery concepts to ease the formulation of the sum-rate maximization optimization problem in the next section.}

\textit{1) Physical layer model:}
\ignore{For FAP-user and CE-D2D user-user channel models, we adopt the SUI model and distance-depend path-loss, where the distance between FAP/CE-D2D user and user is the main factor in modeling the channel.}Let $P_k$ and $Q_n$ denote the transmission power of the $k$-th F-AP and the $n$-th CE-D2D user, respectively. To avoid complexity, we consider fixed transmission power for the CE-D2D users.\ignore{The distance between the $k$-th F-AP  and $u$-th user is denoted by $d_{k,u}$ and the distance between the $n$-th CE-D2D and the $u$-th user is denoted by $D_{n,u}$.} We consider  quasi-static fading channels between the $k$-th F-AP and the $u$-th user and between the $n$-th CE-D2D user and the $u$-th user, which are assumed to
remain fixed in each transmission.  The channel fading gain for the links between $k$-th F-AP and the $u$-th user and between the $n$-th CE-D2D user and the $u$-th user are denoted by $\left|h_{k,u}\right|^2$ and $\left|H_{n,u}\right|^2$, respectively.  These channel gains are summarized in the matrix $\mathbf H$. The signal-to-interference-plus-noise (SINR)  for the links between the $k$-th F-AP and the $u$-th user and the $n$-th CE-D2D user and the $u$-th user are given by, respectively, $
\gamma_{k,u}=\frac{P_k \left|h_{k,u}\right|^2}{\sum_{k' \in \mathcal{K}, k'\neq k}P_{k'} \left|h_{k',u}\right|^2 +N_0}$ and $\gamma_{n,u}=\frac{Q_n \left|H_{n,u}\right|^2}{\sum_{n' \in \mathcal{N}, n'\neq n}Q_n' \left|H_{n',u}\right|^2 +N_0},$ 
where $N_0$ denotes the additive white Gaussian noise variance. 

For F-AP and CE-D2D user transmissions, files are transmitted
via cellular and D2D links, respectively. The data rate of the $u$-th user assigned to the $k$-th F-AP can be expressed as $R_{k,u}=\log_2\left(1+\gamma_{k,u} \right)$ and the data rate of the $u$-th user assigned to the $n$-th CE-D2D can be expressed as $C_{n,u}=\log_2\left(1+\gamma_{n,u} \right)$. To ensure a successful delivery of files to users, the $k$-th F-AP and the $n$-th CE-D2D user can transmit at a rate which is at most equal to the minimum rate of their assigned users, i.e., $R_{k}\leqslant R_{k,u}$ and $C_{n}\leqslant C_{n,u}$, respectively. The set of achievable capacities of all users in all F-APs and CE-D2D users can be represented, respectively, as
$\mathcal{R} = \bigotimes_{(k,u) \in \ \mathcal{\mathcal{K}} \times \mathcal{U}} R_{k,u}$ and $\mathcal{C} = \bigotimes_{(n,u) \in \ \mathcal{\mathcal{N}} \times \mathcal{U}} C_{n,u}$, where the symbol $\bigotimes$ represents the product of the set of the achievable capacities.

\textit{1) Network layer model:}
\ignore{  To maximize the number of successfully received bits per second, i.e., sum-rate, F-APs and CE-D2D users exploit the diversity of users' requests to combine them using XoR operation and transmit encoded files to interested users.Specifically, each F-AP and each CE-D2D user can perform XoR operation on users' requests and send the combined XoRed files to the interested users.After each transmission,
users feedback to the FAPs and neighboring CE-D2D users an acknowledgment for each received file, and accordingly, their \emph{Has}  and \emph{Wants} sets are updated. For convenience, the term ``targeted users" is referred to scheduled users who receive an instantly-decodable transmission.}Let $\kappa_{k}$ and $\mathtt f_{n}$ be the file combinations transmitted from the $k$-th F-AP and $n$-th CE-D2D user, respectively, to the set of  targeted users  $\tau(\kappa_{k})$ and $\mathtt u(\mathtt f_{n})$. Since each user can be scheduled to F-AP $k$ via cellular channel or to  CE-D2D user $n$ via D2D link, we have $\tau(\kappa_{k}) \cap \mathtt u(\mathtt f_{n})=\emptyset$. 
These file combinations are in-fact elements of  the  power set (i.e., XOR combinations of the files) of the available files at the \textit{caches}  of the $k$-th F-AP and the $n$-th CE-DED user. In other terms, $\kappa_{k}\in \mathcal P(\mathcal F_k)$ and $\mathtt f_{n}\in \mathcal P(\mathcal F_n)$. For convenience, the term ``targeted users" is referred to scheduled users who receive an instantly-decodable transmission. A transmission  from  the $k$-th F-AP is instantly decodable at the $u$-th user if it contains the requested file by the $u$-th user and the scheduled transmission rate at the $k$-th F-AP, $R_k$, is no larger than the channel  capacity $R_{k,u}$. Mathematically, $u \in \tau(\kappa_{k})$ will hold if and only if $\{u \in \mathcal{U} \ \big||\kappa_{k} \cap \mathcal{W}_{u}| = 1~\text{and}~ R_{k} \leq R_{k,u}\}$. Similarly, for D2D links, a transmission  from  the $n$-th CE-D2D user is instantly decodable at the $u$-th user if: i) it contains the requested file by the $u$-th user and the scheduled transmission rate at the $n$-th CE-D2D user, $C_n$, is no larger than the channel  capacity $C_{n,u}$ and ii) the $u$-th user is in the coverage zone of the $n$-th CE-D2D user. Mathematically, the set of targeted users by the $n$-th CE-D2D user is expressed as $\mathtt u(\mathtt f_{n})=\{u\in \mathcal U\big||\mathtt f_{n} \cap \mathcal{W}_{u}| = 1~\text{and}$ $ u\in \mathcal{A}_{n} ~\text{and}~ C_n \leq C_{n,u}\}$.

\subsection{Caching Policy and File Delivery}
\ignore{ Our two phases in this work are, namely, (i) caching phase and (ii) file delivery phase. In this subsection, we describe some concepts that ease the formulation of the sum-rate maximization optimization problem in the next section.} 

\textit{1) Caching policy:} Let $\mathbf S$ denote the side information matrix which summarizes the \textit{Has} and \textit{Wants} sets of all the users in a binary $U \times F$  matrix wherein the entry $(u,f)$ represents that the $f$-th file is requested by the $u$-th user. Let $\mathbf C$ denote the file caching matrix where $(f, k)$ represents file $f$ is cached at F-AP $k$ and $(f, n)$ represents file $f$ is cached at CE-D2D user $n$. Owing to their cache sizes, F-APs and CE-D2D users can cache a limited number of files.  If some files are missing in the local caches, they need to be fetched from the CBS via capacity-limited fronthaul links. Therefore, the data rate for fetching the $f$-th file of the $u$-th user from the $k$-th F-AP is $\mathtt R_{k,u}=\min\left(C_{fh}, R_{k,u}\right)$. Accordingly, the set of users whose requests are being missed at the network is denoted by $\mathcal U_m$, i.e., $\mathcal U_m \subset \mathcal U$ and the set of F-APs who can serve users in $\mathcal U_m$ is denoted by $\mathcal K_m$.

\textit{2) File transmission policy:} There are
generally two types of transmission policies, CLNC F-AP transmission via F-RAN cellular link and CLNC CE-D2D transmission via D2D link. Hence, each F-AP $k$ and each CE-D2D user $n$ implement CLNC optimization mechanism that decides their scheduled transmission rates, $R_k$, $C_n$, to send their optimal file combinations $\kappa_k$, $\mathtt f_n$ to set of targeted users $\tau(\kappa_k)$, $\mathtt u(\mathtt f_n)$, respectively. In addition, the F-APs control
the deleterious impact of interference on data rates by employing a power allocation mechanism.\ignore{ As such, users can obtain their requested files with the maximum possible sum-rate.}

\ignore{
\textit{3) User decoding policy:} Targeted users performs an XoR decoding policy with the received file combinations to retrieve their requested files. Consider that the \textit{Wants} sets of user $1$ and user $2$ are $\mathcal W_1=\{f_1\}$, $\mathcal W_2=\{f_2\}$, respectively, and their \textit{Has} sets are $\mathcal H_1=\{f_2\}$, $\mathcal H_2=\{f_1\}$, respectively. Assume that the XoR combination of F-AP $k$ is $\kappa_k=f_1\oplus f_2$, and thus $\kappa_k$ can be decoded by both users since it combines their requested files. User $1$ performs $\kappa_k \oplus f_2=(f_1\oplus f_2)\oplus f_2$ to retrieve $f_1$ and user $2$ performs $\kappa_k \oplus f_1=(f_1\oplus f_2)\oplus f_1$ to retrieve $f_2$.

\textit{4) Performance metric:} We adopt the sum-rate as the performance
metric of interest that measures the total number of successfully received bits per second at the users. For the aforementioned example, assume that the data rates of user $1$ and user $2$ from F-AP $k$ are $2$ bps and $2.5$ bps, respectively. To ensure successful files delivery to both users, F-AP $k$ adopts a minimum transmission rate of $R_k=\min(2, 2.5)=2$ bps and transmits $\kappa_k$ to both users.  Since both users can decode their requested files from $\kappa_k$, the corresponding sum-rate is $2R_k=4$ bits/s.}

\section{Problem Formulation}\label{PF}
The sum-rate maximization optimization problem in F-RAN system with D2D communications involves a joint optimization of an instantaneous-optimization problem for file delivery and a long-term
optimization problem for file placement. \ignore{ Specifically, it involves the NC user scheduling and power
optimization problem for F-AP and CE-D2D user transmissions and the file placement  problem for efficiently caching files at F-APs and CE-D2D users.Consequently, in the first subsection, we formulate the sum-rate maximization optimization problem for file delivery and then employ it for file placement problem formulation in the second subsection.}

\subsection{Problem Formulation for File delivery}
We introduce three binary variables $x_{u,f}$, $y_{f,k}$, $v_{f,n}$ as such $x_{u,f}=1$ if the $u$-th user requests the $f$-th file, and $x_{u,f}=0$ otherwise; $y_{f,k}=1$ if the $f$-th file is stored in the local cache of the $k$-th F-AP, and $y_{f,k}=0$ otherwise; and $v_{n,k}=1$ if the $f$-th file is stored in the local cache of the $n$-th CE-D2D user, and $v_{n,k}=0$ otherwise. Let $l_k$ denote the maximum number of missing files at the $k$-th F-AP that need to fetched from the CBS. The instantaneous-optimization problem to be
solved by the F-APs and the CE-D2D users is formulated as $\mathcal P_1$ given at
the top of the next page.
\begin{table*}
	\vspace*{-0.4cm}
	\begin{normalsize}	
\begin{subequations}
\begin{align} \nonumber \label{eqn:Mschedule11}
& \mathcal{P}_1: \hspace{0.2cm}
\max_{\substack{x_{i,j} \in \{0,1\}\\ R_k, C_n, P_k, \tau_k, \mathtt u_n}} s=\sum _{\substack{k\in \mathcal{K}  }} \sum_{u\in \tau_k }x_{k,u}\min_{u \in \mathcal{\tau}_{k}} R_{k,u}+ \sum _{\substack{n\in \mathcal{N}  }} \sum_{u\in \mathtt u_n }x_{n,u}\min_{u \in \mathtt u_n} R_{n,u}+\sum _{\substack{k \in \mathcal{K}_m\\u\in \tau_k \cap \mathcal{U}_m  }} x_{k,u} \min_{u \in \mathcal{\tau}_{k}} \mathtt R_{k,u}\\
&\rm s.t.
 \begin{cases}  \nonumber
 \hspace{0.2cm} \text{C1:}\hspace{0.2cm} \sum _{f} \sum_{u\in \mathcal U} x_{u,f}(1-y_{f,k})\leqslant l_k,~\forall k \in \mathcal K, \\
\hspace{0.2cm} \text{C2:}\hspace{0.2cm} \tau(\kappa_{k}) \cap \tau(\kappa_{k'})=\emptyset ~ \mbox{ \& }  ~\mathtt u(\mathtt f_{n}) \cap \mathtt u(\mathtt f_{n'})=\emptyset, \forall (k,k')\in \mathcal K, (n,n') \in \mathcal N,\\
\hspace{0.2cm} \text{C3:}\hspace{0.2cm} \tau(\kappa_{k}) \cap \mathtt u(\mathtt f_{n})=\emptyset, \forall k \in \mathcal K, n \in \mathcal N,\\
 \hspace{0.2cm} \text{C4:}\hspace{0.2cm} \kappa_{k}\subseteq \mathcal P(\mathcal{H}_{k}) \mbox{ \& } \mathtt f_{n}\subseteq \mathcal P(\mathcal{H}_{n}), ~\forall k\in \mathcal K, n \in \mathcal N,\\
\hspace{0.2cm}  \text{C5:}\hspace{0.2cm} R_{k}\geq R_{\text{th}} \mbox{ \& } C_{n}\geq R_{\text{th}}, ~\forall k\in \mathcal K, n \in \mathcal N,\\
 \hspace{0.2cm} \text{C6:}\hspace{0.2cm} 0 \leq P_{k} \leq P_{\max}, ~\forall k\in \mathcal{K}.
  \end{cases}
\end{align}
\end{subequations}
	\end{normalsize}
\vspace*{-0.5cm}
\hrulefill
\end{table*}

In $\mathcal P_1$, C1 implies that the number of missing files at each F-AP is limited by $l_k$, due to the capacity-constrained fronthaul links; C2 implies
that the set of scheduled users to all F-APs are disjoint, and similarly, the set of scheduled users to all CE-D2D users are disjoint; C3 makes sure that no user can be scheduled to F-AP and CE-D2D user at the same time instant; C4 ensures that all files to be combined using
XOR operation at all the F-APs and the CE-D2D users are stored in their local caches, respectively; C5 satisfies the minimum transmission rates required to meet the QoS rate requirement $R_{\text{th}}$; and C6 bounds the maximum transmit power of each F-AP. $\mathcal P_1$ is a mixed integer non-linear programming problem and it has NP-hard complexity.

\ignore{$\mathcal P_1$ contains user associations to F-APs and CE-D2D users parameters $x_{k,u}$, $x_{n,u}$, NC scheduling parameters $\mathtt u(\mathtt f_{n}), \tau(\kappa_{k}), \forall k\in \mathcal K, \forall n\in \mathcal N$, power allocations of F-APs  $P_{k},~k\in \mathcal K$, CE-D2D users and their transmission rates.}
 
	
To address the computational intractability of $\mathcal{P}_1$, we consider the following two  clustering strategies, namely, F-AP clustering and CE-D2D user clustering.

\textit{1) F-AP clustering:} Users are grouped into disjoint clusters, where in each cluster there is one F-AP. Let $\Phi$ denote the set of F-AP clusters and expressed as $\Phi=\{\psi_1 ,\ \cdots, \psi_m\}$, $\psi_i \subset \mathcal K$,
$\psi_i \cap \psi_j=\emptyset, \forall i,j \in\{1,\ \cdots,m\}~ \text{and}~ i \neq j$. Within each cluster, CLNC scheme is adopted. To mitigate the inter-cluster interference due to the simultaneous transmissions of F-APs in the clusters, CLNC employs power allocation to efficiently allocate  power levels to the F-APs. As such, a potential
number of users can be targeted with NC combination with improved sum-rate.

\textit{2) CE-D2D user clustering:} The considered realistic scenario of the partially connected D2D networks motivates us to naturally group CE-D2D users and their neighboring users into clusters. Thus, each CE-D2D user and its neighboring users can form a cluster $Z$. Let $\mathbf Z$ denote the set of all disjoint clusters which can be expressed as $\mathbf Z=\{Z_1,\ \cdots, Z_n\}$, $Z_i \subset \mathcal N, \forall i,j \in\{1,\ \cdots,n\}$.

Therefore, the file delivery problem is solved in three steps. At first, we  obtain the aforementioned clusters. Subsequently, we obtain the network-coded file delivery decision in each cluster. Finally, we optimize transmit power allocations of the FAPs to mitigate interference in the system.

\subsection{Problem Formulation for File Placement}
The sum-rate maximization in $\mathcal P_1$ is
not only related to $\mathbf H$ and $\mathbf S$, but also on the file placement matrix $\mathbf C$, and consequently, $\mathcal P_1$ is highly affected by $\mathbf C$. For maximizing the benefit of caching at F-APs and CE-D2D users, the file placement problem taking $\mathcal P_1$ into account have two objectives, namely, (i)  maximize the system
sum-rate and (ii) reduce the cost caused by cache
deployments at F-APs and CE-D2D users and file pushing.
Therefore, the sum-rate maximization optimization problem that includes $\mathcal P_1$ is formulated as $\mathcal P_2$ given at
the top of the next page.
\begin{table*}
	\vspace*{0.12cm}
	\begin{normalsize}	
\begin{subequations}
	\begin{align} \nonumber 
& \mathcal{P}_2: \hspace{0.2cm}
\max_{\mathbf C} \Phi=\omega\mathbb{E}_{\mathbf H, \mathbf S}\left[s(\mathtt S^*(\mathbf C))\right]- \mu \left(\sum _{\substack{f}} \sum_{k} y_{f,k}+ \sum _{\substack{f}} \sum_{n} v_{f,n}\right)\\
& \rm s.t.
 \begin{cases}  \nonumber
 \hspace{0.2cm} \text{C1}-\text{C6},\\
 \hspace{0.2cm} \text{C7:}\hspace{0.2cm} \sum _{f} \sum_{u\in \mathcal U} x_{u,f}y_{f,k}\leqslant \mu F,~\forall k \in \mathcal K, \\
\hspace{0.2cm} \text{C8:}\hspace{0.2cm} \sum _{f} \sum_{u\in \mathcal U} x_{u,f}v_{f,n}\leqslant \mu F,~ \forall n \in \mathcal N,\\
\hspace{0.2cm} \text{C9:} \hspace{0.2cm} y_{f,k}\in \{0,1\} ~\& ~v_{f,n}\in \{0,1\}.
  \end{cases}
\end{align}
\end{subequations}
	\end{normalsize}
\vspace*{-0.5cm}
\hrulefill
\end{table*}

In $\mathcal P_2$, $\mathtt S^*$ is the NC and user scheduling of the F-APs and CE-D2D users,  $\omega$ is the weight factor that represents the benefit
of unit long-term sum-rate,  and $\mu$ is the cost of caching a file. C7 and C8 make sure that the number of users' requested files that are cached by each F-AP and each CE-D2D user, respectively, are limited, due to the constrained cache size.  We can readily show that problem $\mathcal{P}_2$ is NP-hard and intractable. Specifically, $\mathcal{P}_2$ is a two-level problem formulation, namely, (i) upper-level caching problem and (ii) lower-level file delivery problem $\mathcal{P}_1$ (i.e., NC, power allocation, resource scheduling). We emphasize that $\mathcal{P}_2$ is computationally intractable since both $\mathcal{P}_1$ and the caching optimization problem are NP-hard.


The caching decision $\mathbf C$ depends on the resource allocation solution of $\mathcal P_1$, and unfortunately, the solution of $\mathcal{P}_1$ also depends on $\mathbf C$. Therefore, the overall optimization is very complex and can only be solved optimally via exahaustive. However, this is not suitable for real-time applications. Moreover, the decision of $\mathtt S^*$ and $\mathbf C$ occurs in two different time scales, namely, (i) the objective of $\mathcal P_1$ is to instantly improve the sum-rate and (ii) the objective of caching is to maximize the long term rate. Distributed RL is a suitable platform to solve problem $\mathcal P_2$ whose goal is maximizing long term sum-rate, and its main idea is that agents  do not have any prior information about the system (i.e., the resource allocation decision). Hence, agents react with environment, learn the instantaneous reward (i.e., instantaneous rate), and update their strategy. As such, agents can achieve benefit in the long run. To this end, capitalizing the distributed RL and game theory, we propose an effective and efficient framework to solve $\mathcal{P}_2$ with the reduced computational complexity.

\subsection{A Game Perspective of Problem $\mathcal P_1$ and Problem $\mathcal P_2$}
\textit{1) Game description of $\mathcal P_1$:} The F-AP and CE-D2D user clustering process can be modeled as a coalition formation game (CFG) \cite{CFG}, which facilitates the development of a fully distributed and low-complexity
algorithm. Specifically, our objective is to develop a distributed framework that models the collaborations among the F-APs and CE-D2D users of F-RAN system. For this objective, we use CFG because it studies the cooperative behavior of F-APs and CE-D2D users in maximizing the sum-rate.\ignore{ The F-APs and CE-D2D users find their alliances
to form cooperative coalitions so that they maximize the sum-rate.}
\ignore{Coalitional game has been used to
model cooperative behavior among network nodes \cite{46, 47,48,49,50}. Prior network coding works with coalition games aimed
to maximize sum-rate, see the works in \cite{51}, \cite{52}. In particular, the authors of \cite{52} showed the potential of NC in maximizing the sum-rate in network coding-aided D2D communication. Moreover, some prior works analyzed the problem of content dissemination when NC is enabled, e.g., \cite{54}, \cite{55}.}Our studied game is formally expressed as $\mathcal G_{F}\left(\mathcal U, (s_u)_{u\in \mathcal U}, (\succ_u)_{u\in \mathcal U}\right)$. In $\mathcal G_F(.)$, $\mathcal U$ is the set of followers that represent the set of users, $(s_u)_{u\in \mathcal U}$ the utility function user $u$ and it is the rate of that user, and  $(\succ_u)_{u\in \mathcal U}$ is the preference of user $u$. Therefore, we can write the utility function of the $u$-th user and its preference $(\succ_u)$ in F-AP clustering and CE-D2D user clustering, respectively, as
\begin{equation}
s_u=R_{k},
\end{equation}
\begin{align} \label{eq4}
\psi \succ_u \psi(u) \Leftrightarrow
\begin{cases}
|\mathcal W_u \cap \mathcal{H}(\psi)| = 1  \mbox{ \& } s_u(\Phi_n) > s_u(\Phi_o),\\
\sum_{u'\in \psi} s_{u'}(\Phi_n) > \sum_{u'\in \psi(u)} s_{u'}(\Phi_o),\\
\sum_f\sum_{u'\in \psi}x_{u',f}(1-y_{f,k})\leq l_k, \forall k\in \mathcal K,
\end{cases}
\end{align}
and 
\begin{align} \label{eq5}
Z \succ_u Z(u) \Leftrightarrow
\begin{cases}
|\mathcal W_u \cap \mathcal{H}(Z)| = 1 \mbox{ \& }  s_u(\mathbf Z_n) > s_u(\mathbf Z_o) ,\\
\sum_{u'\in Z} s_{u'}(\mathbf Z_n) > \sum_{u'\in Z(u)} s_{u'}(\mathbf Z_o),\\
\sum_f\sum_{u'\in Z}x_{u',n}(1-v_{f,n})\leq l_n, \forall n\in \mathcal N.
\end{cases}
\end{align}
In \eref{eq4}, $\Phi_n$, $\Phi_o$ are the old and new F-AP clustering after user $u$ switches from $\Phi(u) \in \Psi_o$ to  $\Phi \in \Psi_n$; and in \eref{eq5}, $\mathbf Z_n$, $\mathbf Z_o$ are the old and new CE-D2D user clustering after user $u$ switches from $Z(u) \in \mathbf Z_o$ to  $Z \in \mathbf Z_n$.
In \eref{eq4} and \eref{eq5}, user $u$ prefers another cluster
over its current cluster if and only if: (i) it can receive its requested file and its data rate for being in that new cluster is strictly improved; (ii) the sum data rate of users involved in the new cluster of F-AP/CE-D2D user should be increased; and (iii) after user $u$
joins the new cluster, the fronthaul capacity constraint of the
FAP/CE-D2D user in  that cluster is still satisfied.

\textit{2) Game description of $\mathcal P_2$:} The two-level problems in $\mathcal P_2$, as mentioned before, are coupled and have sequential decisions, and interestingly, the interaction between these decisions can be modeled as a Stackelberg game. In Stackelberg game, 
the F-APs and CE-D2D users are leaders and the users are followers  \cite{Game}. The equilibrium of such a game is expressed as $\mathcal S_E(\mathbf C^*, \Psi^*_{\mathbf C^*})$. In $\mathcal S_E(.)$, $\Psi^*_{\mathbf C^*}$ is the F-AP and CE-D2D user clustering when the best strategy of caching is $\mathbf C^*$. 
$\mathcal S_E(.)$ is a Stackelberg equilibrium (SE) if and only if: $s(\mathbf C^*, \Psi^*_{\mathbf C^*}) \geq  s(\mathbf C, \Psi^*_{\mathbf C}), \psi^*(t)_{u} \succ_u \psi^*_{i}, Z^*(t)_{u} \succ_u Z^*_{i}(t), \forall u, \forall t, \forall \Phi^*_i(t) \in \Psi^*(t)/ \{\Phi^*_u(t)\}, Z^*_i(t) \in \Psi^*(t)/ \{Z^*_u(t)\}$,
where $\psi^*_u(t) \in \Phi^*(t)$ and $Z^*_u(t) \in \mathbf Z^*(t)$  are the clusters to
which users are assigned given the caching strategy is $\mathbf C^*$.

\ignore{Once $\mathcal S_E(.)$ is achieved, each user has no incentive to deviate from its cluster formed in any transmission
interval if the strategies of the caching optimization at
F-APs and CE-D2D user are fixed. Meanwhile, the multi-agent leaders have no
incentive to adjust cache resources.}

\section{Sum-Rate Maximization: Joint Approach}\label{JA}
\ignore{This section first develops an innovative decentralized CLNC-CF switch algorithm
for F-AP and CE-D2D user clustering, under a fixed strategy of file caching. Then, taking the behavior
of F-APs and CE-D2D users into account, a  multi-agent RL algorithm is developed to achieve a local optimal strategy for file caching.}

\subsection{ CLNC-CF Switch Algorithm}\label{CLNC-CFG}
We consider two kinds of coalitions in our CFG becasue we have two kinds of clustering in our system, F-AP clustering and CE-D2D clustering. In our CFG, players switch coalitions in order to optimize overall
system sum-rate. The players choose whether to switch coalitions
or not according to a pre-defined preference
order $\succ$. Such preference order in our CFG is
related to the system sum-rate, and each player should
be able to obtain it by relying only on local network
information. A framework of the proposed decentralized CLNC-CF switch algorithm is presented as follows.

\textbf{Phase I:} In this phase, we describe the switching process of players among F-AP and CE-D2D user coalitions.

\textit{Switching among F-APs:}
Given the initialized random coalition for F-APs, this step optimizes the selection of user-F-AP assignment through several successive switch operations between F-AP coalitions. Specifically, players switch their coalitions based on preference order $\succ$ that is related to the sum-rate. The switch operations are implemented by checking the switch possibilities of player $u$ to each coalition $\psi$. Initially, we have $\Phi_\text{ini}$. Thus, a player $u$ can decide to switch its current coalition $\psi(u)$ to a new coalition $\psi$ if the following holds: (i) the data rate of player $u$ is strictly improved without  affecting the utilities of all the remaining players in $\psi$ and (ii) the requested file by player $u$ is cached at the associated F-AP in $\psi$ or the fronthaul capacity link of associated F-AP in $\psi$ is still satisfied after joining player $u$.  Therefore, player $u$ should switch coalition $\psi(u)$ and join coalition $\psi$, and accordingly, $\Phi_\text{ini}$ is updated.

\textit{Switching among CE-D2D users:}
Given the initialized random coalition for CE-D2D users, this step optimizes the selection of user-CE-D2D user assignment using many switch operations. Specifically, players switch their coalitions based on preference order $\succ$ that is related to the sum-rate. The switch operations are implemented by checking the switch possibilities of player $u$ to each coalition $Z$. Initially, we have $\mathbf Z_\text{ini}$. Thus, a player $u$ can decide to switch its current coalition $Z(u)$ to a new coalition $Z$ if the following holds: (i) the data rate of player $u$ is strictly improved without  affecting the utilities of all the remaining players in $Z$ and (ii) the requested file by player $u$ is cached at the associated CE-D2D user in $Z$ or the fronthaul capacity link of associated CE-D2D user in $Z$ is still satisfied after joining player $u$.  Hence, player $u$ should switch coalition $Z(u)$ and join coalition $Z$, and consequently, $\mathbf Z_\text{ini}$ is updated.

\textit{Remark 1: Since the system constraints must be maintained, switching coalitions
cannot take place if any of the constraints is violated.
It can be seen that this switching process is completely
decentralized and only local network information is
required by the users. Specifically, player $u$ only needs to know the data rates of its cellular and D2D links.}

\textbf{Phase II:} In this phase, we develop a CLNC scheme that considers NC user scheduling and power level control of the resource allocation of F-APs and CE-D2D users. Given the resulting coalition formation process by the first phase of Algorithm 1, $\mathcal P_1$ can be rewritten as
\begin{equation*}
\begin{split}
\mathcal{P}_3: \hspace{0.2cm}
&\max_{\substack{R_k, C_n, P_k, \tau_k, \mathtt u_n}} s=\sum _{\substack{k\in \mathcal{K}  }} \sum_{u\in \tau_k }\min_{u \in \mathcal{\tau}_{k}} R_{k,u}\\ & +\sum _{\substack{n\in \mathcal{N}  }} \sum_{u\in \mathtt u_n }\min_{u \in \mathtt u_n} R_{n,u}+\sum _{\substack{k \in \mathcal{K}_m\\u\in \tau_k \cap \mathcal{U}_m  }} \min_{u \in \mathcal{\tau}_{k}} \mathtt R_{k,u}\\
{\rm s.t.\ }
& \text{C4-C6}.
\end{split}
\end{equation*}
$\mathcal P_3$ is a  NC user scheduling and non-convex power allocation problem which is NP-hard and computationally intractable. 
In this work, capitalizing the graph theory, we propose an iterative method to solve $\mathcal P_3$ with the reduced computational complexity.  For the given transmit power
allocation of each F-AP and CE-D2D user, $\mathcal P_3$ is solved using maximum weight clique (MWC) search method. Afterward, for the resulting NC user scheduling, the power allocation problem is solved numerically. These two steps are iterated until convergence.

\textit{Graph description:} Let $\mathcal G_k(\mathcal V_k,\mathcal E_k)$ and $\mathcal G_n(\mathcal V_n,\mathcal E_n)$ denote the distributed-graphs of the $k$-th F-AP and the $n$-th CE-D2D user, wherein $\mathcal V$ and $\mathcal E$ are their corresponding vertices and edges, respectively. Each F-AP and each CE-D2D user generate their vertices based on their cached files and the scheduled users in their coalitions. A vertex $v \in \mathcal V_k$ in $\mathcal G_k$ is generated for each user $u \in \psi_k$, its requested file $\mathcal W_u$, 
and for each achievable rate for that user $r\in\mathcal R_u=\{R\in \mathcal R| R\leq R_{k,u}\}$. In a similar way, a vertex $v \in \mathcal V_n$ in $\mathcal G_n$ is generated for each user $u \in Z_n$, its requested file $\mathcal W_u$, 
and for each achievable rate for that user $r\in\mathcal C_u=\{C\in \mathcal C| C\leq C_{n,u}\}$. Two vertices in $\mathcal G_k$ or in $\mathcal G_n$ (representing the same transmitter\footnote{For different transmitters (F-APs or CE-D2D users), vertices always represent different users as this is already guaranteed from the clustering process in the first phase of Algorithm \ref{Algorithm1-a}. Hence, such vertices are connected.}) are connected if they satisfy the instantly decodable NC constraints and their corresponding rates are equal \cite{29}. Consequently,
$\bigcup\limits^K_{k=1}\mathcal{V}_{k}$ and $\bigcup\limits^N_{n=1}\mathcal{V}_{n}$ give the entire graph $\mathcal G$ and its corresponding vertices are $\mathcal{V} = \bigcup\limits_{l=\{1,\cdots, K, 1, \cdots, N\}}\mathcal{V}_{l}$.
The weight of each vertex $v$ is the rate of the represented user as follows
\begin{equation}\label{v}
w(v)=
\begin{cases}
R_{k,u} ~~\text{if $v \in \mathcal V_k$}, \forall k\in \mathcal K,\\
C_{n,u} ~~\text{if $v \in \mathcal V_n$}, \forall n\in \mathcal N,\\
\min\left(C_{fh}, R_{k,u}\right), ~~\text{if $v \in \mathcal V_k$}, \forall k\in \mathcal K_m.
\end{cases}
\end{equation}
Recall, $w(v)=\min\left(C_{fh}, R_{k,u}\right)$ represents the case of missing requested files at F-AP $k$ that have to be fetched from the CBS via fronthaul link with capacity $C_{fh}$.

\textit{MWC description:}
With the designed distributed graph $\mathcal G$,  $\mathcal P_3$ is similar to
MWC problems in several aspects. In MWC, two vertices must be connected in the graph,
and similarly, in $\mathcal P_3$, two different users
must be scheduled to two different transmitters. Moreover, our objective in $\mathcal P_3$ is to maximize
the total sum-rate, and similarly, the
goal of MWC is to maximize the weight of all vertices. Consequently, we have the following lemma.

\textit{\textbf{Lemma 1:} Using the distributed graph $\mathcal G$, $\mathcal P_3$ can be equivalently transformed to the problem of determining the MWC in $\mathcal G$.}

\begin{proof} 
	Assuming that $\mathbf{X}^*=\{v_{1},v_{2}, \cdots, \, v_{|\mathbf X|}\},~ \forall v \in \mathcal G$ is the MWC. Let $\mathbf X$ is the set of all possible cliques in $\mathcal G$. For each $v \in \mathbf X^*$ that is associated with an association of (user, file, transmitter, and rate), the weight $w(v)$ in \eref{v} is the utility that the induced user in $v$ can receive. Hence, the weight of the MWC $\mathbf{X}^*$  is mainly the objective function of $\mathcal P_3$ and can be written as $ w(\mathbf{X}^*)= \sum\limits _{v\in \mathbf{X}^*}w(v)=\sum _{\substack{k\in \mathcal{K}  }} \sum_{u\in \tau_k }\min_{u \in \mathcal{\tau}_{k}} R_{k,u}+ \sum _{\substack{n\in \mathcal{N}  }} \sum_{u\in \mathtt u_n }\min_{u \in \mathtt u_n} R_{n,u}+\sum _{\substack{k \in \mathcal{K}_m\\u\in \tau_k \cap \mathcal{U}_m  }} \min_{u \in \mathcal{\tau}_{k}} \mathtt R_{k,u}$  Since two vertices that represent different users are scheduled to different transmitters are adjacent, constraints C2, C3 hold. Moreover, all the vertices are indeed generated based on the cached files of F-APs and CE-D2D users, and accordingly, constraint C4 holds. Finally, the set of targeted users and file combinations is obtained by combining the vertices of the MWC $\mathbf{X^*}_l$ corresponding to $\mathcal{G}_{l}, \forall l\in \mathcal K \cup \mathcal N$. Therefore, $\mathbf{X}^*=\bigcup^{K+N}_{l=1} \mathbf{X^*}_l$. 
	
\end{proof}

\textit{F-AP power control optimization:}
For the stable coalition  of F-APs that results in NC and F-AP user scheduling, obtained from solving $\mathcal P_3$,  the power allocation problem is written as
\begin{equation*}
\begin{split}
&\mathcal{P}_4: 
\max_{\substack{P_k}} \sum _{\substack{k\in \mathcal K }} |\tau_t| * \min_{u \in \mathcal{\tau}_{k}} R_{k,u}+ \sum _{\substack{s\in \mathcal K_m  }} |\tau_u| * \min_{u \in \mathcal{\tau}_{s}} \mathtt R_{s,u},\\
&{\rm s.t.\ }
0 \leq P_{k} \leq P_{\max},
\end{split}
\end{equation*}
To  solve $\mathcal{P}_4$ effectively, we derive optimal power allocations
to maximize sum-rate for a given NC user scheduling. 
 To this end, we introduce the following lemma.

\textit{\textbf{Lemma 2:}
	Given the resulting NC and user-F-AP schedule by Algorithm 1, a converged power
	allocation is obtained by updating power $\{P_k\}_{k\in \mathcal K}$ at the $(t + 1)$-th iteration, based on the following power update manipulation
	\begin{eqnarray}
	\label{Power_update}
	P_k^{t+1}=\left[\frac{|\tau_k|*\frac{\gamma_{k,\hat{u}}}{1+\gamma_{k,\hat{u}}}}{\sum^K_{\substack{{l=1, l\neq k}}} \left(|\tau_l|*\frac{(\gamma_{l,\hat{v}})^2}{1+\gamma_{l,\hat{v}}}\right)\frac{h_{k,\hat{v}}}{{P}_l h_{l,\hat{v}}}}\right]_{0}^{P_{\max}},
	\end{eqnarray}
	where  $\hat{u}=\arg \min_{u \in \mathcal{\tau}_{k}}R_{k,u}$,  $\hat{v}=\arg \min_{v \in \mathcal{\tau}_{l}}R_{l,v}$, $\forall k, l \in \mathcal{K}$ and $k \neq l$;  $\gamma_{k,\hat{u}}=\frac{P_{k}^{(t)} h_{k,\hat{u}}}{1+\sum_{m=1,m\neq k}^KP_{m}^{(t)} \hat{h}_{m,\hat{u}}}$; and 
	$P_k^{(t)}$ is the transmit power of the $k$-th F-AP at the $t$-th iteration.}

\proof 	The proof is analogous to \cite[Proposition 2]{Saif_TVT}, and is omitted due to the space limitation.

The two-phase CLNC-CF switch algorithm is distributively executed at each iteration index and summarized in Algorithm \ref{Algorithm1-a}.
\begin{algorithm}[t!]
	\caption{CLNC-CF Switch Algorithm for File Delivery}
	\label{Algorithm1-a}
	\begin{algorithmic}[1]
		\State \textbf{Input:} $\mathcal N$, $\mathcal K$, $\mathcal H_{u}$, $\mathcal W_{u}$, $\forall u\in \mathcal U$, $\forall n\in \mathcal N$.
		\State \textbf{Initialize:}  F-AP and CE-D2D user random clustering $\Psi_\text{ini}=\{\Phi_\text{ini}, \mathbf Z_\text{ini}\}=\{\psi_1, \ \cdots, \psi_m, Z_1, \cdots , Z_n\}$, $\psi_i \cap \psi_j=\emptyset$, $Z_k \cap Z_l=\emptyset$, $\psi_i \cap Z_l=\emptyset$, $\forall i\neq j, k\neq l$, $\Psi_\tau=\Psi_\text{ini}$.
		\State \textbf{Phase I:} \textbf{Coalition switch process}
		\Repeat 
		\State	\textit{Switching among F-APs:}
		\For{$u\in \Phi_\text{ini}$} 
		
		\For{$\psi_i \in \Psi_\text{ini}, \forall i=\{1,2, \cdots, m\}$}
	
		\If{$\psi \succ_u \psi(u)$}
		
		\State Update $\Phi_\text{ini}\leftarrow \Phi_\text{ini}\backslash \{\psi(u), \psi\} \cup \{   \psi(u)\backslash \{u\}, \psi \cup \{u\}\}$.
		
		\EndIf
		
		\EndFor
		\EndFor	
		
		\State	\textit{Switching among CE-D2D users:}
	    \For{$u\in \mathbf Z_\text{ini}$} 
		
		\For{$Z_i \in \mathbf Z_\text{ini}, \forall i=\{1,2, \cdots, n\}$}
	
		\If{$Z \succ_u Z(u)$}
		
	\State Update $\mathbf Z_\text{ini}\leftarrow \mathbf Z_\text{ini}\backslash \{Z(u), Z\} \cup \{  Z(u)\backslash \{u\}, Z \cup \{u\}\}$.
		\EndIf
	 \EndFor
		\EndFor	
	   \Until{No further switch operation}
		\State \textbf{Phase II:} \textbf{Solve $\mathcal P_3$ using CLNC:}
		\State \textit{MWC search method:}
		\State Initialize $\mathbf X^*  = \emptyset$ and  $\mathcal{G}(\mathbf X^*) \leftarrow \mathcal{G}$.
		\State $\forall v \in\mathcal{G}$, calculate $w(v)$ as in \eref{v}.
		
		\While{$\mathcal{G}(\mathbf X^*) \neq \emptyset$}
		\State Choose the maximum weight $v^*=\arg\max_{v\in \mathcal{G}(\mathbf X^*)} \{w(v)\}$.
		\State Set $\mathbf X^* \leftarrow \mathbf X^* \cup v^*$ and $\mathcal{G}(\mathbf X^*) \leftarrow \mathcal{G}(v^*)$.
		\EndWhile
		\State Output $\mathbf X^*$.
		\State \textit{F-AP power allocation optimization:}
		\State Using the F-AP-user association in $\mathbf X^*$, solve $\mathcal P_4$ to compute the optimal power allocations.
	\end{algorithmic}
\end{algorithm}

\subsection{Multi-Agent Reinforcement Learning}\label{RL}
After obtaining the resulting coalitions formation by Algorithm \ref{Algorithm1-a} for a given file placement strategy, our remaining task to obtain a SE for the formulated Stackelberg game is to near-optimally place files at the local caches, taking the behaviors of F-APs and CE-D2D users into account. Achieving a SE requires a global strategy of all F-APs' and CE-D2D users' actions in the system. Clearly, such strategy is unavailable to the F-APs and CE-D2D users
in practice, and hence, achieving an SE solution is challenging. To address this challenge, we leverage multi-agent RL (MARL) that allows each F-AP and each CE-D2D users to construct its strategy without having global information.

Let $(a_{i,j})_{i=\{1,\cdots, D\}, j\in \{1,2\}}$ is the $j$-th action of the $i$-th virtual agent, where  $D$ is the number of virtual agents in the network, i.e.,  $D=FK+FN$. The action set of $D$ virtual agents in the $t$-th index is defined as $ \underline{\bm a}_t= \{(a_{1,1}, a_{1,2}), (a_{2,1}, a_{1,2}), \cdots, (a_{D,1}, a_{D,2})\}$, where each action represents a possible cache matrix $\mathbf C$. Specifically, the action space of the $i$-th agent is $(a_{i,1}, a_{i,2})$ and when $i = (k-1)F+f$, taking action $a_{i,1}$ means F-AP $k$ caches file $f$, while taking $a_{i,2}$ means  F-AP $k$ does not cache file $f$. Similarly, when $i = (n-1)F+f$, taking action $a_{i,1}$ means CE-D2D user $n$ caches file $f$, while taking $a_{i,2}$ means  CE-D2D user $n$ does not cache file $f$.  

Let $\Gamma$ denote the history profile, $\Gamma= \{\mathbf H_1, \mathbf S_1\}, \{\mathbf H_2, \mathbf S_2\}, \cdots, \{\mathbf H_T, \mathbf S_T\}$, where $\{\mathbf H_1, \mathbf S_1\}\in \Gamma$ is the tuple of channel gain matrix and side information matrix in the $t$-th iteration index. By exploiting this history network data, the virtual agents learn and achieve  sub-optimal caching strategy. Assuming that 
action $a_i^t$ is chosen, Algorithm \ref{Algorithm1-a} is
executed, and a feedback signal is provided to the virtual agent to guide the learning process, The feedback signal for the $i$-th virtual agent, denoted by $r^{a_t}_i$, is given by
\begin{equation}\label{11}
r^{a_t}_i=ws(a^t_i, \mathbf H_t, \mathbf S_t)-\mu \sum_f \sum_k y_{f,k}(a^t_i).
\end{equation}
In \eref{11}, $s(a^t_i, \mathbf H_t, \mathbf S_t)$ is the system sum-rate in the $t$-th iteration index defined in $\mathcal P_2$ and $y_{f,k}(a^t_i)$ represents the caching decision associated with file $f$ at transmitter $k$ under action $a^t_i$.
	
After computing $r^{a_t}_i$, each virtual agent $i$ implements two RL procedures to estimate the expected utility
received from Algorithm 1 and to properly selects an action probability, respectively. Let $\underline{\bm x}^t_{i,j}=\left[(x^t_{i,1},x^t_{i,2}), \cdots, (x^t_{D,1},x^t_{D,2})\right]_{i\in \{1,\cdots, D\}, j\in\{1,2\}}$ be the vector of estimated utilities and $\underline{\bm \rho}^t_{i,j}=\left[(\rho^t_{i,1},\rho^t_{i,2}), \cdots, (\rho^t_{D,1},\rho^t_{D,2})\right]_{i\in \{1,\cdots, D\}, j\in\{1,2\}}$ be the vector of action probability of all agents, respectively, at the $t$-th th iteration index. Consequently, to strike a balance between exploration and exploitation, we introduce a variable $\sigma$, and thus, we have $\beta(x^t_{i,j})=\frac{\exp(\sigma x^t_{i,j})}{\exp(\sigma x^t_{i,1})+\exp(\sigma x^t_{i,2})}$. The utility estimation and action probability update
equations are expressed as, respectively,
\begin{align}\label{12}
x^{t+1}_{i,j}=\alpha_t\mathbb I_{\{{a^t_i=a_{i,j}}\}}\left(r^{a_t}_i-x^{t}_{i,j}\right)+x^{t}_{i,j}, \forall i\in \{1,\cdots, D\},
\end{align}

\begin{align}\label{13}
\rho_{i,j}^{t+1}=\rho_{i,j}^{t}+\lambda_t\left(\beta(x^t_{i,j})-\rho_{i,j}^{t})\right), \forall i\in \{1,\cdots, D\}. 
\end{align}
In \eref{12} and \eref{13}, $\lambda_t$ and $\alpha_t$ are learning rates and $\mathbb I_{\{a^t_i=a_{i,j}\}}=1$ if action $a_{i,j}$ is taken in the $t$-th iteration index
interval $t$ and $0$ otherwise. Similar to \cite{Peng_ML}, $a^t_i$ follows the same definition that all virtual agents share the same feedback signal, and, $\lambda_t$ and $\alpha_t$ should satisfy the following conditions $\sum_{t\geq 1} \lambda_t=\infty,~~ \sum_{t \geq 1} \lambda^2_t \leq \infty, \sum_{t\geq 1} \alpha_t=\infty,~~ \sum_{t \geq 1} \alpha^2_t \leq \infty,\lim_{t \rightarrow \infty} \frac{\lambda_t}{\alpha_t}=0$.
To calculate \eref{12} and \eref{13}, the $i$-th agent initializes a utility
vector and an equal probability action vector, given by, $x^1_{1,j}=\left[(0,0)\right]_{j\in \{1,2\}}$ and $\rho^1_{1,j}=\left[(0.5,0.5)\right]_{j\in \{1,2\}}, \forall i$, respectively. With the increase in the $t$-th index, the
parameter $\sigma$ in \eref{13} continuously grows, and correspondingly, each virtual agent intends
to explore the two actions and choose the action associated with the highest expected utility. To perform this, we consider  the well-known $\epsilon$-greedy policy that allows the agents to choose exploration and exploitation with probability $\epsilon$ and $1-\epsilon$, respectively. To summarize, each virtual agent iteratively repeats the following three steps, (i) \textit{action selection}, (ii) \textit{learning calculation}, and (iii) \textit{update step}  that are presented in Algorithm  \ref{Algorithm1}.

 \begin{algorithm}[t!]
	\caption{Multi-agent RL based Caching
		Decision}
	\label{Algorithm1}
	\begin{algorithmic}[1]
		\State \textbf{Initialize:} Initializes a virtual environment
		based on history network data  $\Gamma= \{\mathbf H_1, \mathbf S_1\}, \{\mathbf H_2, \mathbf S_2\}, \cdots, \{\mathbf H_T, \mathbf S_T\}$.
		\State Create virtual agents $D=KF+NF$.
		\State Initialize $t=1$, $\lambda_t$, and $\alpha_t$.
		\State Set $x^t_{i,j}=\left[(0,0)\right]_{j\in \{1,2\}}$ and $ \rho^t_{i,j}=\left[(0.5,0.5)\right]_{j\in \{1,2\}}, \forall i$.
		
		\Repeat 
		\State \textit{Action selection:}
		\For{$i=1 ~\text{to} ~D$} 
		\State Generate a random number $l_i$.

		\If{$l_i< \rho^t_{i,1}$}
		
		\State Virtual agent $i$ takes action $a_{i,1}$.
		
	    \Else 
	    \State Virtual agent $i$ takes action $a_{i,2}$.
		\EndIf 	
		\EndFor
		\State Under $\underline{\bm a}_t$, $\mathbf H_t$, and $\mathbf S_t$, Algorithm 1 is
		executed to output $s(a^t_i, \mathbf H_t, \mathbf S_t)$.
        \State \textit{Learning and update:}
	    \For{$i=1 ~\text{to} ~D$} 
	    \State Calculate the learning rate $r^{a_t}_i$ by using \eref{11}.
	    \State Update $x^t_{i,j\in \{1,2\}}$ and $ \rho^t_{i,j\in \{1,2\}}$ by using \eref{12} and \eref{13}, respectively.
	    \EndFor

		\Until{$t=T$}
	\end{algorithmic}
\end{algorithm}
\ignore{
The  aforementioned developed two-algorithms of solving $\mathcal P_2$ that are explained in \sref{CLNC-CFG} and \sref{RL}, respectively,  are iterated until $T$ iteration. Algorithm \ref{Algorithm1} provides a sub-optimal, yet efficient, solution to the sum-rate maximization optimization problem $\mathcal{P}_2$ while requiring low computational complexity as will be explained in the next section. To make the overall process of Algorithm \ref{Algorithm1} more intuitive, Fig. \ref{fig2} is drawn.

\begin{figure}[t!]
	\centering
	\includegraphics[width=0.65\linewidth]{fig2nn.pdf}
	\caption{An illustration of the CLNC-RL approach.}
	\label{fig2}
\end{figure}}

\section{The Properties of the Proposed Approach}\label{PP}
In this section, we analyze the properties of the two-algorithms proposed approach, in
terms of convergence, optimality and complexity.

\subsection{CLNC-CF Switch Algorithm}
\textit{1) Convergence:} The switching process of users in stage $1$ and stage $2$ is converged becasue there is an infinite number of F-APs and CE-D2D users and thus, infinite number of switch operations. It can be observed that
the achieved data rate by user $u$ in any coalition formation of CE-D2D users is not influenced by the coalition
formation of F-APs outside the coalition to which it belongs. Therefore, when the clustering of F-APs or CE-D2D users is updated based on the preference order, the system
sum-rate will strictly increase. Otherwise, users would not switch. Since sum-rate is bounded,
after finite repeats of Algorithm \ref{Algorithm1-a}'s steps, the clustering of F-APs and CE-D2D users will finally not
change. Hence, Algorithm \ref{Algorithm1-a} will converge.

\textit{2) Stability:} 
The stability of the final clustering depends on the existence of a Nash-stable
solution for our CF game \cite{E_H}. Let $\Psi_{fin}=\{\Phi_{fin}, \mathbf Z_{fin}\}$ denote
the final clustering obtained by Algorithm \ref{Algorithm1-a}. Once Algorithm \ref{Algorithm1-a} converges to $\Psi_{fin}$, it can be readily concluded that $\psi \succ_u \psi(u)$ for
any user $u$ and any $\psi \in \Phi_{fin}/\{\psi(u)\}$, and similarly, $Z \succ_u Z(u)$ for
any user $u$ and any $Z \in \mathbf Z_{fin}/\{Z(u)\}$. Such a stable state is Nash-stable \cite{E_H}.

\ignore{
\textit{2) Optimality:} The local optimality and
Nash stability relationship is discussed in the following theorem. 

\textit{\textbf{Theorem 1:}
Any local optimal F-AP and CE-D2D user coalition formation must be Nash-stable,
while a Nash-stable coalition formation may not be local optimal.}

\begin{proof}
Given a local optimal clustering formation, switching any user
from its current cluster to another cluster is not guarantee to contribute to a strict increase of system sum-rate. This is becasue some users may have same utility in different clusters. Consider that there exists
a local optimal solution that is not Nash-stable. In this case, some users are willing to join a different coalition, which will strictly improve their sum-rate according to the
preference order definition. Hence, contradiction occurs and it is concluded
that any local optimal clustering must be Nash stable. In the contrary, since a Nash-stable solution is achieved based on
the preference order that requires a strict improvement of user’s
own performance, it is still possible to further increase the system sum-rate by switching users from their current clusters to different clusters under a Nash-stable partition. Thus, a Nash-stable clustering formation may not be local optimal.
\end{proof}}

\textit{3) Complexity:} At each iteration, Algorithm \ref{Algorithm1-a} needs to check
at most $KU$ and $NU$ preference orders for switching process in phase I, and consequently, the computational complexities
of phase I execution is $\mathcal O(KU+NU)$. For $L$ iterations until convergence, the computational complexity is $\mathcal O(L(KU+NU))\approx \mathcal O(LKN)$, where $N>K$. Next, we analyze the computational complexity of phase II. For finding the NC user scheduling, each transmitter (i.e., F-AP and CE-D2D user) needs to build its distributed graph and find the corresponding MWC. The computational complexity of constructing the graph and finding its MWC of transmitter $t$ is bounded by $\mathcal O(|\mathcal F_t|^2U^2)$ and $\mathcal O(U|\mathcal F_t|^2)$ \cite{13}.  This gives a total complexity of $\mathcal O((N+K)(|\mathcal F_t|^2U^2+U|\mathcal F_t|^2))\approx \mathcal O(N|\mathcal F_t|^2U^2K)$ that is associated with finding the NC combinations in the system \cite{13}. For solving the power allocation problem in $\mathcal P_4$, Algorithm \ref{Algorithm1-a} needs $C_p=\mathcal O\left(|\mathtt{S_1}| \times |\mathtt{S_2}| \times \cdots |\mathtt S_{N+K}|\right)$ where $\mathtt{S_1}$ represents the NC user scheduling of the first transmitter.  Therefore, the overall worst-case computational complexity of Algorithm \ref{Algorithm1-a} is $\mathcal O(LNU+C_p+N|\mathcal F_t|^2U^2K)$, which can be approximated to a polynomial computational complexity of $\mathcal O(N|\mathcal F_t|^2U^2K)$.

\subsection{MARL based Caching Algorithm}
\ignore{In this subsection, we prove the convergence and near-optimality of Algorithm \ref{Algorithm1}. Moreover,
the overall computational complexity is presented as well.}

\textit{1) Convergence:} At each iteration, Algorithm \ref{Algorithm1} updates the virtual agents’ probability of
selecting the actions. Thereby, at the convergence of Algorithm \ref{Algorithm1}, the action selection probability
of the virtual agent become stable. Such a fact is demonstrated by the following theorem.\\
\textit{\textbf{Theorem 1:}
	The $i$-th virtual agent's action selection probability is converged as $\lim_{t \to \infty}\rho_{i,j}^{(t)} \to \rho_{i,j}^* $, $\forall i \in \{1,2,\cdots,D\}$ and $j \in \{1,2\}$. Here
	\begin{equation}
	\label{prob_Converge}
	\rho_{i,j}^*=\frac{\exp\left(\sigma \overline{r}_{i,j}\right)}{\sum_{j=1}^2 \exp\left(\sigma \overline{r}_{i,j}\right)}
	\end{equation}
	where $\overline{r}_{i,j}$ is the $i$-th virtual agent's expected reward from the $j$-th action.}

\begin{proof}
	According to \cite[Theorem 4]{Mehdi_learning}, under the condition of $\lim_{t \to \infty} \frac{\alpha_t}{\lambda_t}=0$,  we obtain that the following two conditions are satisfied. \textbf{c1:} $\lim_{t \to \infty}\left|x_{i,j}^{t}-\overline{r}_{i,j}\right|=0$, $\forall i,j$; and \textbf{c2:} As $t \to \infty$, the update equation for the action selection probability of the virtual agent's action converges to the following ordinary differential equation (ODE).
%
	\begin{equation}
	\label{ODE}
	\dot{\rho}_{i,j}^{t}=\beta(x_{i,j}^{t})-\rho_{i,j}^{t}
	\end{equation}
Note that, as $\lim_{t \to \infty} \frac{\alpha_t}{\lambda_t}=0$,  the action selection probabilities are updated relatively slowly compared to the estimated utility.  On the other hand, as per \textbf{c1}, the estimated utilities are converged as $t \to \infty$. Substituting the converged value of the estimated utility in \eqref{ODE} and using the fact that at the stationary point of \eqref{ODE}, $\dot{\rho}_{i,j}^{t}=0$, the  value of the stationary action selection probability is obtained as
	\begin{equation}
	\label{converge_prob}
	\rho_{i,j}^*=\beta( \overline{r}_{i,j}).
	\end{equation}
	Next, we  justify that \eqref{converge_prob} provides the converged action selection probability for the virtual agents.  Based on \cite[eq. (2)]{RL}, the optimal action selection probability is obtained  by maximizing the expected reward of the virtual agents augmented by the entropy function. Accordingly, for the $i$-th virtual agent, $\forall i \in \{1,2,\cdots,D\}$, we obtain
	\begin{equation}
	\label{converge_prob_2}
	\begin{split}
&	\left\{\rho_{i,1}^*, \rho_{i,2}^*\right\}=\arg \max_{\rho_{i,1} \geq 0, \rho_{i,2}  \geq 0} \sum_{j=1}^2 \rho_{i,j}\overline{r}_{i,j}\\ & +\frac{1}{\sigma}\sum_{j=1}^2 \rho_{i,j}\ln\left(\frac{1}{\rho_{i,j}}\right),~ \text{s.t.} \quad \sum_{j=1}^2 \rho_{i,j}=1. 
	\end{split}
	\end{equation}
	In \eqref{converge_prob_2}, the factor $\frac{1}{\sigma}$ controls the relative importance of maximizing the total expected reward and entropy functions. Applying the Lagrangian optimization technique, we can readily obtain
	\begin{equation}
	\label{converge_prob_3}
	\rho_{i,j}^*=\frac{\exp\left(\sigma \overline{r}_{i,j}\right)}{\sum_{j=1}^2 \exp\left(\sigma \overline{r}_{i,j}\right)}\triangleq\beta( \overline{r}_{i,j}).
	\end{equation}
	Evidently, as $t \to \infty$, the action selection probability of virtual agents  becomes stable, and such a stable action selection probability also coincides with the optimal action selection probability for the virtual agents. Therefore,  Algorithm \ref{Algorithm1} converges as the number of iterations is increased.
\end{proof}

\textit{2) Optimality:} The optimality of Algorithm \ref{Algorithm1} is confirmed from the following theorem.

\textit{\textbf{Theorem 2:}
Algorithm \ref{Algorithm1} provides a near-optimal solution to $\mathcal P_2$ optimization problem.}
\begin{proof}
	Since \eqref{converge_prob} is obtained at the maximal point of \eqref{converge_prob_2}, we can readily justify
	\begin{equation}
	\label{optimality_1}
	\begin{split}
	&\overline{r}_{i,j}(\rho_{i,j}^*, \bm{\rho}_{-i})+\frac{1}{\sigma} \rho_{i,j}^*\ln\left(\frac{1}{\rho_{i,j}^*}\right)
	\geq \overline{r}_{i,j}(\rho_{i,j}, \bm{\rho}_{-i}) \\ &+\frac{1}{\sigma} \rho_{i,j}\ln\left(\frac{1}{\rho_{i,j}}\right)
\implies \overline{r}_{i,j}(\rho_{i,j}, \bm{\rho}_{-i})-\overline{r}_{i,j}(\rho_{i,j}^*, \bm{\rho}_{-i}) \\ & \leq \frac{1}{\sigma} \left(\rho_{i,j}^*\ln\left(\frac{1}{\rho_{i,j}^*}\right)-\rho_{i,j}\ln\left(\frac{1}{\rho_{i,j}}\right)\right), \forall j
	\end{split}
	\end{equation}
	where $\bm{\rho}_{-i}$	 is the action selection probability of all the virtual agents except the $i$-th virtual agent. Using the upper bound of the entropy function, we can justify $\frac{1}{\sigma} \left(\rho_{i,j}^*\ln\left(\frac{1}{\rho_{i,j}^*}\right)-\rho_{i,j}\ln\left(\frac{1}{\rho_{i,j}}\right)\right) \leq \frac{1}{\sigma} \ln 2 $. Therefore we obtain
	\begin{equation}
	\label{optimality_2}
	\begin{split}
	\overline{r}_{i,j}(\rho_{i,j}, \bm{\rho}_{-i})-\overline{r}_{i,j}(\rho_{i,j}^*, \bm{\rho}_{-i}) \leq \frac{1}{\sigma} \ln 2\triangleq \varepsilon.
	\end{split}
	\end{equation}
	Note that  $\sigma \to \infty$,  $\overline{r}_{i,j}(\rho_{i,j}, \bm{\rho}_{-i}) \leq \overline{r}_{i,j}(\rho_{i,j}^*, \bm{\rho}_{-i})$ is satisfied. In addition, as $\sigma \to \infty$, $\rho_{i,j^*}=1$ where $j^*=\arg\max_{j \in \{1,2\}} \overline{r}_{i,j}$ and $\rho_{i,j \neq j^*}=0$.  Later, we demonstrate that $\sigma$ is an increasing function of the iteration index $t$. As a result, as $t \to \infty$, on the one hand, the action selection of the virtual agents becomes deterministic, i.e., each agent adheres to the action having the maximum expected payoff. On the other hand,  with such an action selection strategy, the expected reward of each agent becomes larger than any given action selection strategy.  Consequently, as $t \to \infty$, Algorithm \ref{Algorithm1} guarantees the  optimal convergence of each virtual agent's achievable reward.   Moreover, we can readily that for the deterministic action selection of the virtual agents, the statistical expected value of \eref{11}  and objective function of $\mathcal P_2$ optimization problem are same. Consequently, as $t \to \infty$, Algorithm \ref{Algorithm1} converges to the joint caching decisions that provides a higher objective value of $\mathcal P_2$ optimization problem than any other joint caching decisions. In other words,  Algorithm \ref{Algorithm1} can converge to the optimal solution to $\mathcal P_2$ optimization problem as $t \to \infty$. 
	
	However, in practice, we can consider a finite number of  iterations for Algorithm \ref{Algorithm1}, and accordingly, from \eqref{optimality_2}, the virtual agents may have certain incentive to deviate from the converged action selection strategy. Nevertheless, eq. \eqref{optimality_2} depicts that with a finite number of iterations, the virtual agents cannot increase their expected payoff more than $\varepsilon$ by deviating from the converged action selection strategy, and the value of $\varepsilon$ decreases as the number of iterations is increased. Hence, we conclude that Algorithm \ref{Algorithm1} provides a near-optimal solution to $\mathcal P_2$ optimization problem. 
\end{proof}

\textit{3) Complexity:} Since each agent needs to make one action, the computational complexity associated
with selecting all the actions in Algorithm \ref{Algorithm1} is $\mathcal O(D)$. Further, the update equations in \eref{12} and \eref{13} involve fixed number of operations, thus the complexity for each virtual agent is $\mathcal O(1)$. Essentially, these corresponding complexities are negligible as compared to the complexity of Algorithm \ref{Algorithm1-a} that is one step in Algorithm \ref{Algorithm1}. Therefore,
the overall worst-case computational complexity of Algorithm \ref{Algorithm1} is \ref{Algorithm1-a}, which is $\mathcal O(TN|\mathcal F_t|^2U^2K)$.

\textit{4) Designing Parameter $\sigma$:} Algorithm \ref{Algorithm1} can
achieve a local optimal caching strategy. However, this result is established on the convergence
when $t \rightarrow \infty$, where we should collect infinite amount of history network data. For limited data, the parameter $\sigma$ should be selected carefully to achieve a good performance. To ease the analysis, we
first proof that the sum of action selection probabilities for each
virtual agent is always equal to $1$.

\textit{\textbf{Theorem 3:}
For each virtual agent $i$, if $\sum_j \rho^1_{i,j}=1$, the update equation in \eref{13} guarantees that $\sum_j \rho^t_{i,j}=1$ with $t\geq 2$.}

\begin{proof}
From \eref{13}, we have $
\sum_j \rho^{t+1}_{i,j}=\sum_j \rho^{t}_{i,j}+\lambda_t \sum_j\left(\beta(x^t_{i,j}-\rho^{t}_{i,j})\right)=\sum_j \rho^{t}_{i,j}+\lambda_t \left(1-\sum_j \rho^{t}_{i,j}\right)$. Thus, it can be noted that $\sum_j \rho^1_{i,j}=1 \Rightarrow \sum_j \rho^2_{i,j}=1 ...,$ and thus $\sum_j \rho^t_{i,j}=1$ with $t\geq 2$.
\end{proof}
The following theorem  analyzes the impact of $\sigma$ on the strategy evolution.

\textit{\textbf{Theorem 4:}
For virtual agent $i$, consider that action $a_{i,1}$
has a larger utility estimation than action $a_{i,2}$, and consequently, $\beta(x^t_{i,1})$
monotonously increases with $\sigma$.}

\begin{proof}
We take the derivative of $\beta(x^t_{i,1})$ with respect to $\sigma$ as follows
\begin{align}\label{13nn}
\frac{d\beta(x^t_{i,1})}{d\sigma}=\beta( x^t_{i,1})\frac{\exp(\sigma x^t_{i,2})(x^t_{i,1}- x^t_{i,2})}{\exp(\sigma x^t_{i,1})+\exp(\sigma x^t_{i,2})}.
\end{align}
Since $x^t_{i,1} > x^t_{i,2}$, we have $\frac{d\beta( x^t_{i,1})}{d\sigma}>0$. Hence, when $\sigma$ becomes larger,
$\beta(x^t_{i,1})$ will increase, and the probability of selecting action $a_{i,1}$
will have a higher chance to increase as in \eref{13}. Consequently, the probability of selecting
action $a_{i,2}$ decreases as in Theorem $3$.
\end{proof}

A practical way of setting $\sigma$
is $\sigma_t=\frac{t}{b}$ \cite{Peng_ML}, \cite{60}, where $b$ is a constant. This setting adjusts the value of $\sigma$ with iteration index $t$. Particularly, $\sigma$ is initially set to a small value to ease the full estimation of utility associated with each action. When $t$ increases, $\sigma$ increases, and the virtual agent selects the
action with the largest utility estimation, and this
improve its own performance.

\section{Numerical Results}\label{NR}
In this section, the effectiveness of the proposed CLNC-CF-RL approach is demonstrated, and corresponding numerical results are conducted.

\subsection{Simulation Setting and Comparison Schemes}
\textit{1) Simulation setting:} We consider an D2D-aided F-RAN where F-APs and CE-D2D users have fixed locations and users are distributed randomly within a hexagonal cell of radius  $1500$m. Unless otherwise stated, we set the radius of the CE-D2D users' transmission range $\mathtt R$ to $500$m and the numbers of F-APs $K$ and CE-D2D users is set to $3$, $6$, respectively.  In addition, each user requests one file in each transmission by following Zipf distribution, and the probability of file $f$ to be requested is given by $\frac{f^\gamma}{\sum_{j=1}^{F}j^\gamma}$, where $\gamma$ is the Zipf parameter that governs the popularity
distribution skewness which is equal to $0.5$. To further emphasize on  the discrepancies among files popularity, each user requests a file with a probability given in the range $[0-0.25]$. The channel model of both F-RAN and D2D communications follow the standard path-loss model, which consists of three components: 1) path-loss of $128.1+37.6\log_{10}(\text{dis.[km]})$ for F-RAN transmissions and path-loss of $148 + 40 \log_{10}(\text{dis.[km]})$ for D2D communications; 2) log-normal shadowing with $4$dB standard deviation; and 3) Rayleigh channel fading with zero-mean and unit variance. The cellular and D2D channels are assumed to be perfectly estimated. The noise power and the maximum’ F-AP and CE-D2D user power are assumed to be $-174$ dBm/Hz and $P_\text{max}=Q=-42.60$ dBm/Hz, respectively. The link bandwidth is $10$ MHz.  Unless otherwise stated, we set $l_k=5, \forall k$  and $C_{fh}=30$ Mbps. For the
update equations in \eref{12} and \eref{13}, the learning parameters are set to $\sigma=\frac{t}{10^5}$, $\alpha_t=\frac{1}{(1+t)^{0.6}}$, $\lambda_t=\frac{1}{(1+t)^{0.7}}$. For the learning rate $r^{a_t}_i$ of each agent $i$ in \eref{11}, we set $\omega=1$ and $\mu=0.8$. To use MARL to learn caching
strategy, $1000$ tuples of $\{\mathbf H,\mathbf S\}$ are generated and utilized
to construct history network environment, and in these history data generations, the locations of F-APs, CE-D2D users, and users
are fixed. 

\textit{2) Comparison schemes:} First, to assess the performance of our proposed CLNC-RL approach with different thresholds ($R_{\text{th}1}=0.5$ and $R_{\text{th}2}=5$), we simulate various scenarios with different number of users and number of files. These thresholds represent  the minimum transmission rates required for QoS. The performances of our proposed solution for $R_\text{th1}$ and $R_\text{th2}$ are shown in solid red line and dash red line, respectively. For the sake of comparison, our proposed schemes are
compared with the following baseline schemes.
\begin{itemize}
	\item \textbf{Optimal Uncoded:} The transmission strategy in this scheme is
	performed irrespective of the available information at the
	network layer, i.e., prior download files. The user-F-AP/CE-D2D user
	association in this scheme is proposed in \cite{38}.
	\item \textbf{Classical IDNC:} For F-AP and CE-D2D user transmissions, this scheme focuses on network layer optimization, where the coding decisions depends solely on the file combinations. 
	\item \textbf{RA-IDNC:} This scheme was studied in \cite{25}.
\end{itemize}

 \begin{figure}[t!]
	\centering
	\begin{minipage}{0.494\textwidth}
		\centering
		\includegraphics[width=0.85\textwidth]{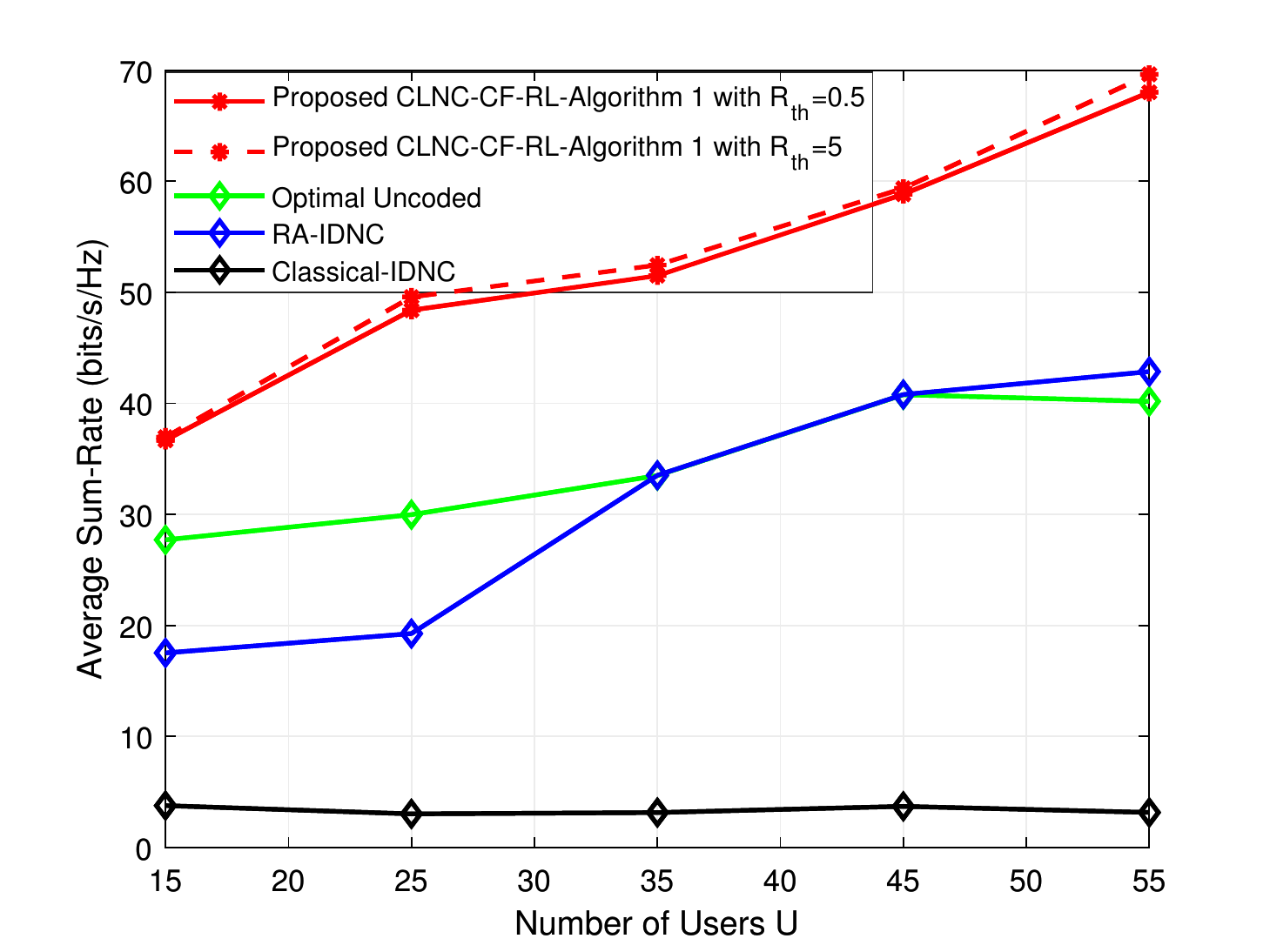} 
		\caption{Sum-rate in vs.  number of users $U$ for a fixed caching with cache size of $0.5F$ and $30$ files.}
		\label{fig3}
	\end{minipage}\hfill
	\begin{minipage}{0.494\textwidth}
		\centering
		\includegraphics[width=0.85\textwidth]{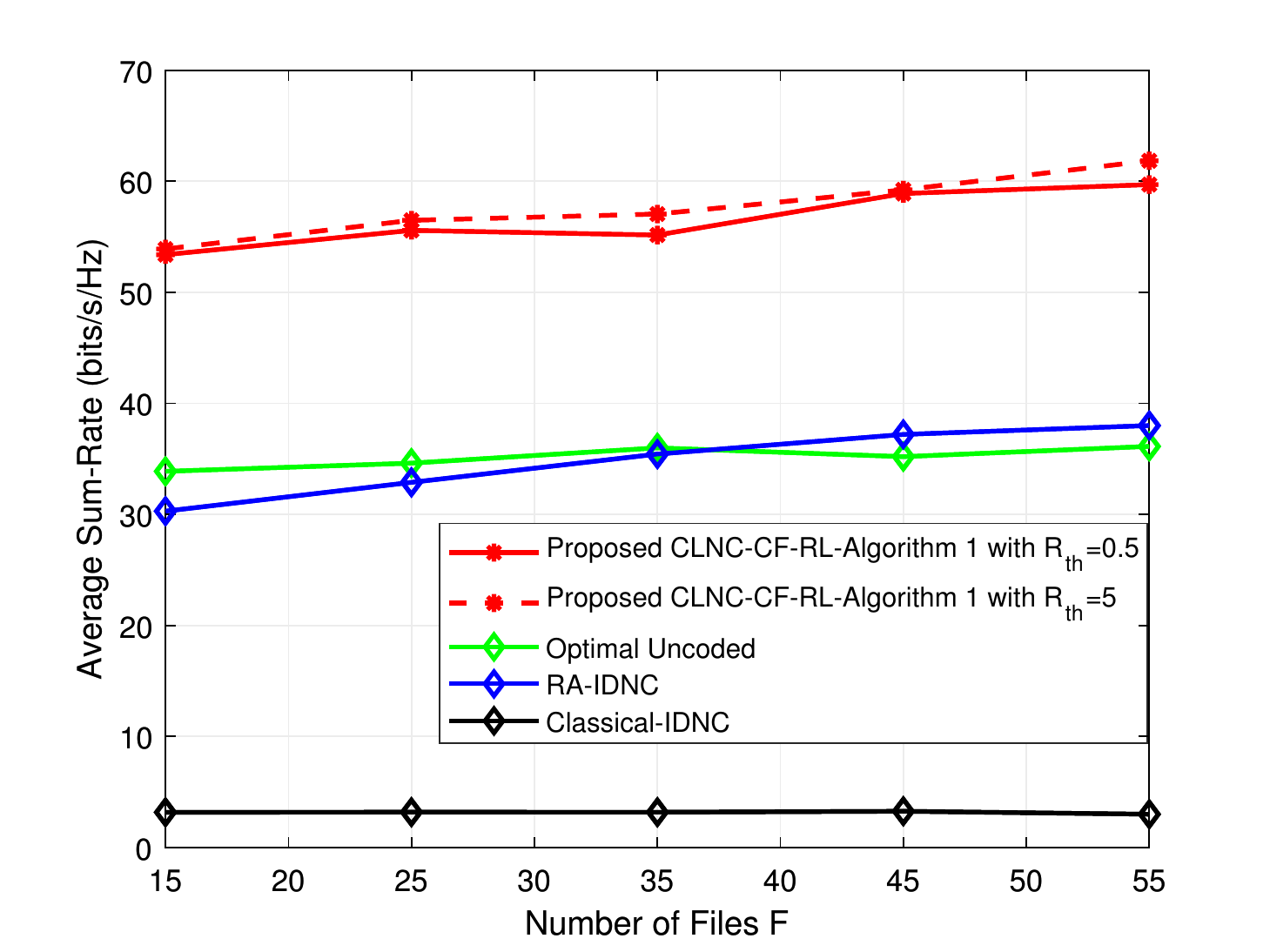} 
		\caption{Sum-rate vs.  number of files $F$ for a fixed caching with cache size of $0.5F$ and $40$ users.}
		\label{fig4}
	\end{minipage}\hfill
\end{figure} 

Second, to illustrate the superior performance of the MARL algorithm, we compare it with other baseline
schemes.
\begin{itemize}
	\item \textbf{All Caching:} In this scheme, each F-AP and each CE-D2D user cache all the files requested by the users.
	\item \textbf{No Caching:}  In this scheme, there is no popular file	cached at F-APs and CE-D2D users.
	\item \textbf{$0.5$ Caching:}  In this scheme, F-APs and CE-D2D users cache files with probability of $0.5$. 
	\item \textbf{Distributed Q-learning:} This scheme was proposed in \cite{Q_RL}. 
\end{itemize}

\subsection{Simulation Discussions}

\begin{figure}[t!]
	\centering
	\begin{minipage}{0.494\textwidth}
		\centering
		\includegraphics[width=0.85\textwidth]{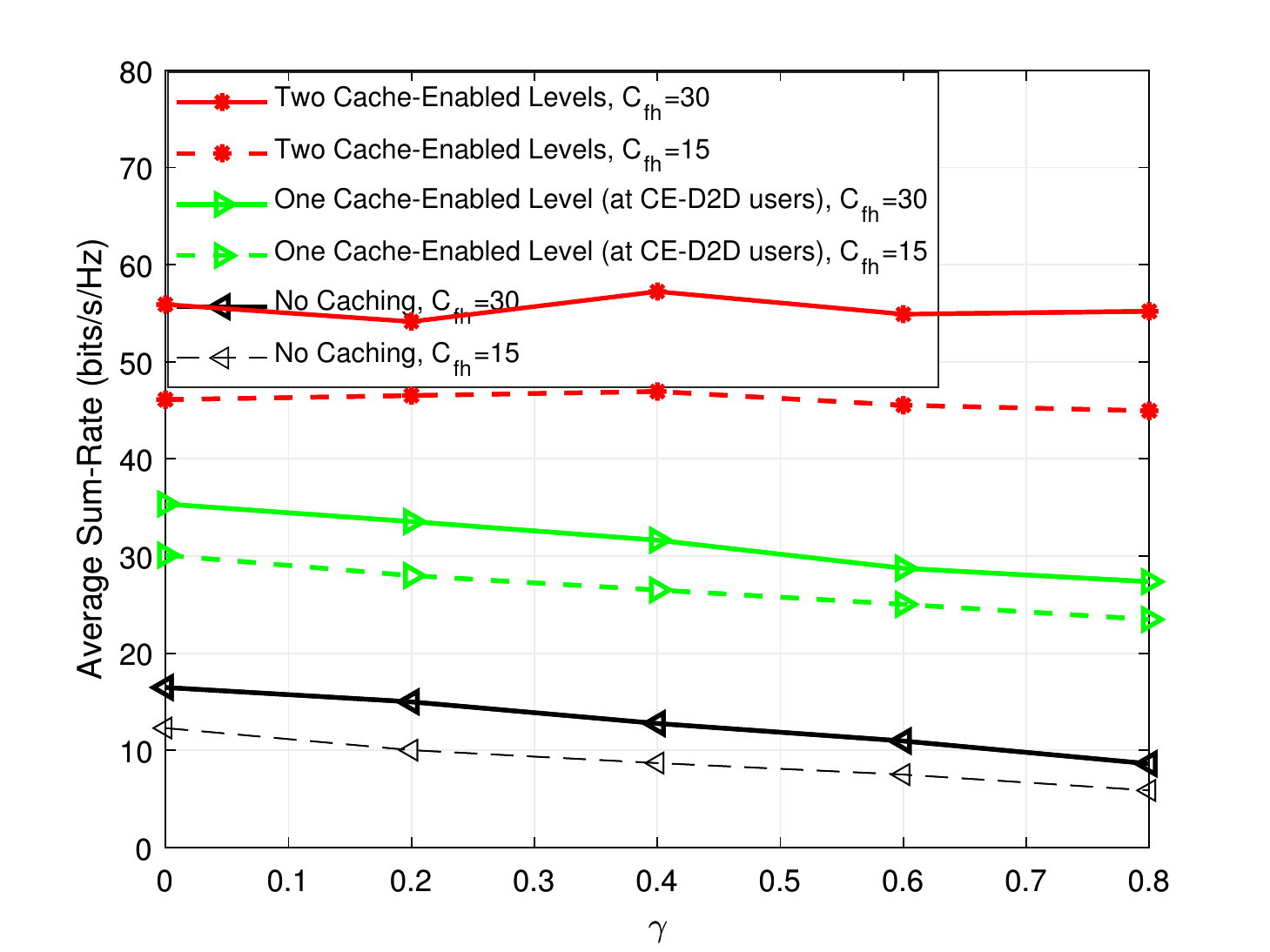} 
		\caption{Sum-rate vs. $\gamma$ with different cache-enabled levels.}
		\label{fig5}
	\end{minipage}\hfill
	\begin{minipage}{0.494\textwidth}
		\centering
		\includegraphics[width=0.85\textwidth]{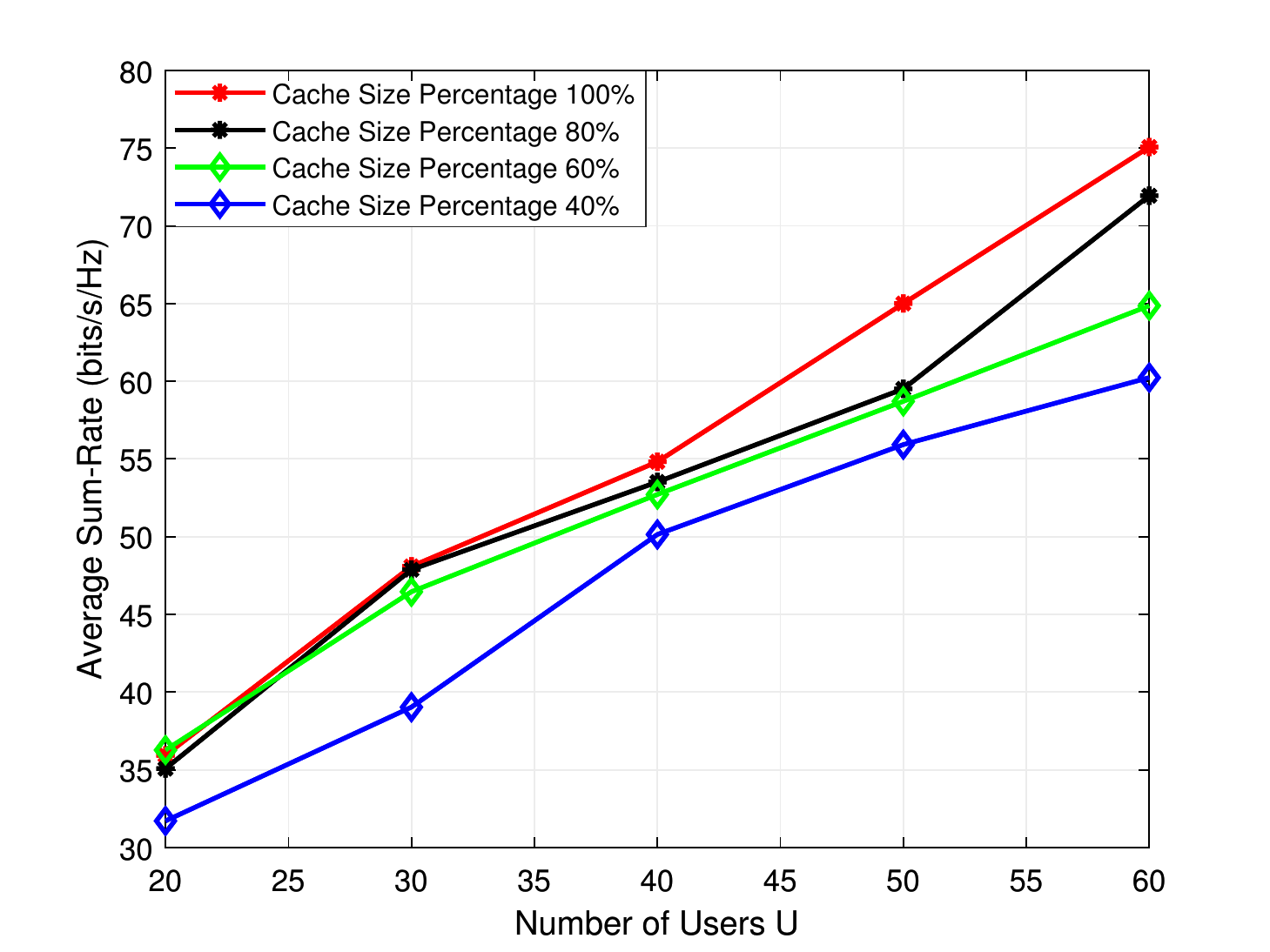} 
		\caption{Sum-rate vs. number of users with different cache sizes.}
		\label{fig6}
	\end{minipage}\hfill
\end{figure}

\textit{1) Simulation of Algorithm \ref{Algorithm1-a}:} In Figs. \ref{fig3} and \ref{fig4}, we present  average sum-rate vs. number of users $U$ for $30$ files and vs. number of files $F$ for $40$ users, respectively\footnote{Each presented value in Fig. \ref{fig3}, \ref{fig4}, \ref{fig5}, and \ref{fig6} is calculated by averaging sum-rate over $1000$ realizations of $\{\mathbf H,\mathbf S\}$. For changing the network topology, the locations of users are randomly generated in each realization.}. Fig. \ref{fig3} shows the impact of multiplexing many users to the F-APs and CE-D2D users using NC on the physical layer performance, and Fig. \ref{fig4} considers different sizes of popular contents. We can see from Figs. \ref{fig3} and \ref{fig4} that the  proposed CLNC-CF  algorithm offers an improved performance in terms of sum-rate as compared to contemporary coded and uncoded schemes. These improved performances are becasue our proposed scheme judiciously schedules users, adopts the transmission rate of each F-APs and optimizes the transmission power of each F-AP, and selects potential users for transmitting coded files over D2D links. Specifically, the classical IDNC scheme suffers from scheduling many users to the FAPs and the CE-D2D users by adopting their transmission rates to the minimum rates of all their scheduled users. Optimal uncoded scheme schedules users to the F-APs and the CE-D2D users based on their strong channel qualities, but it suffers from scheduling few users that is equal to the number of F-APs and CE-D2D users in the network. On the other hand, RA-IDNC scheme offers an improved performance compared to uncoded and classical IDNC schemes since it effectively balances between the number of scheduled users and the transmission rate of F-APs/CE-D2D users. However, selecting one transmission rate (the minimum rate) for all the F-APs and CE-D2D users degrades the sum-rate performance of the RA-IDNC scheme. This is a clear limitation of the RA-IDNC scheme as it does not fully exploit the typical variable channel qualities of the different F-APs and CE-D2D users to their scheduled users. Our proposed CLNC-CF-RL scheme fully utilize the FAPs and CE-D2D users' resources to choose their own transmission rates, file combinations, and scheduled users. Consequently, a better performance of our proposed schemes compared to the RA-IDNC scheme is achieved. Moreover, the joint scheme optimizes the employed rates using power control on each F-AP. Thus, it works better than our proposed coordinated scheme. It is worth remarking that the increase in the number of files in Fig. \ref{fig4} leads to a slight increase in the sum-rate for all coded schemes. This is due to the fact that the number of non-feasible edges between vertices increases as the number of files increases. This results in a smaller IDNC opportunities to combine files, thus leading to a slight improvement in terms of the sum-rate of all NC schemes.

In Figs. \ref{fig5} and \ref{fig6}, we study the impacts of fronthaul capacity and cache size on sum-rate performance of our proposed scheme with different cache-levels, respectively, under various values of $\gamma$. From Fig. \ref{fig5}, it can be seen that a
smaller $\gamma$ leads to a higher sum-rate and larger $\gamma$ leads to a reduction in sum-rate performance. For smaller $\gamma$, most of users will request popular files, and
hence their requests have higher probabilities to be combined using NC and locally met by F-APs and CE-D2D users. On the other hand, when $\gamma$ is high, few files will be requested by users, thus few IDNC possibilities for the F-APs and CE-D2D users.
Furthermore, under a fixed $\gamma$, a higher $C_{fh}$ also contributes for enhancing system performance since users request will be downloaded by a rate of $\min(C_{fh}, R)$.
However, the gain brought by increasing $C_{fh}$ becomes less
for a larger $\gamma$ since more users' requests can be satisfied
locally from the caches of F-APs and CE-D2D users. Consequently, the need for fronthaul capacity is less
stringent. From Fig. \ref{fig6}, we note that a larger cache size makes F-APs and CE-D2D users have more chances to cooperate and thus, system performance is raised.
The improved sum-rate performances in Figs. \ref{fig5} and \ref{fig6} demonstrate  the pronounced role of
our proposed CLNC-CF algorithm with two-level caching in mitigating the fronthaul congestion of dense networks.

\begin{figure}[t!]
	\centering
	\begin{minipage}{0.494\textwidth}
		\centering
		\includegraphics[width=0.85\textwidth]{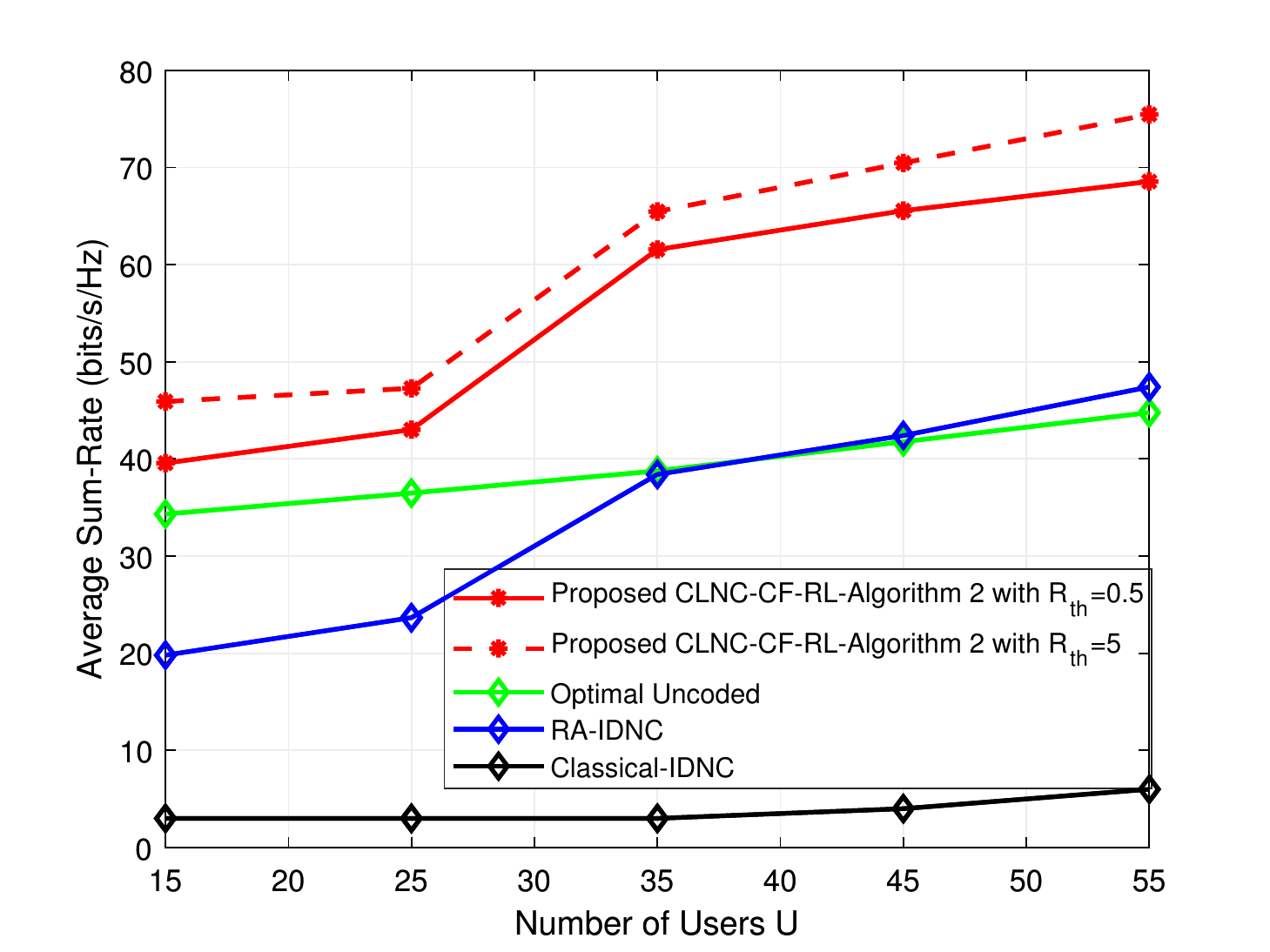} 
		\caption{Sum-rate vs.  number of users $U$ with an optimized caching policy using MARL for $30$ files.}
		\label{fig7}
	\end{minipage}\hfill
	\begin{minipage}{0.494\textwidth}
		\centering
		\includegraphics[width=0.85\textwidth]{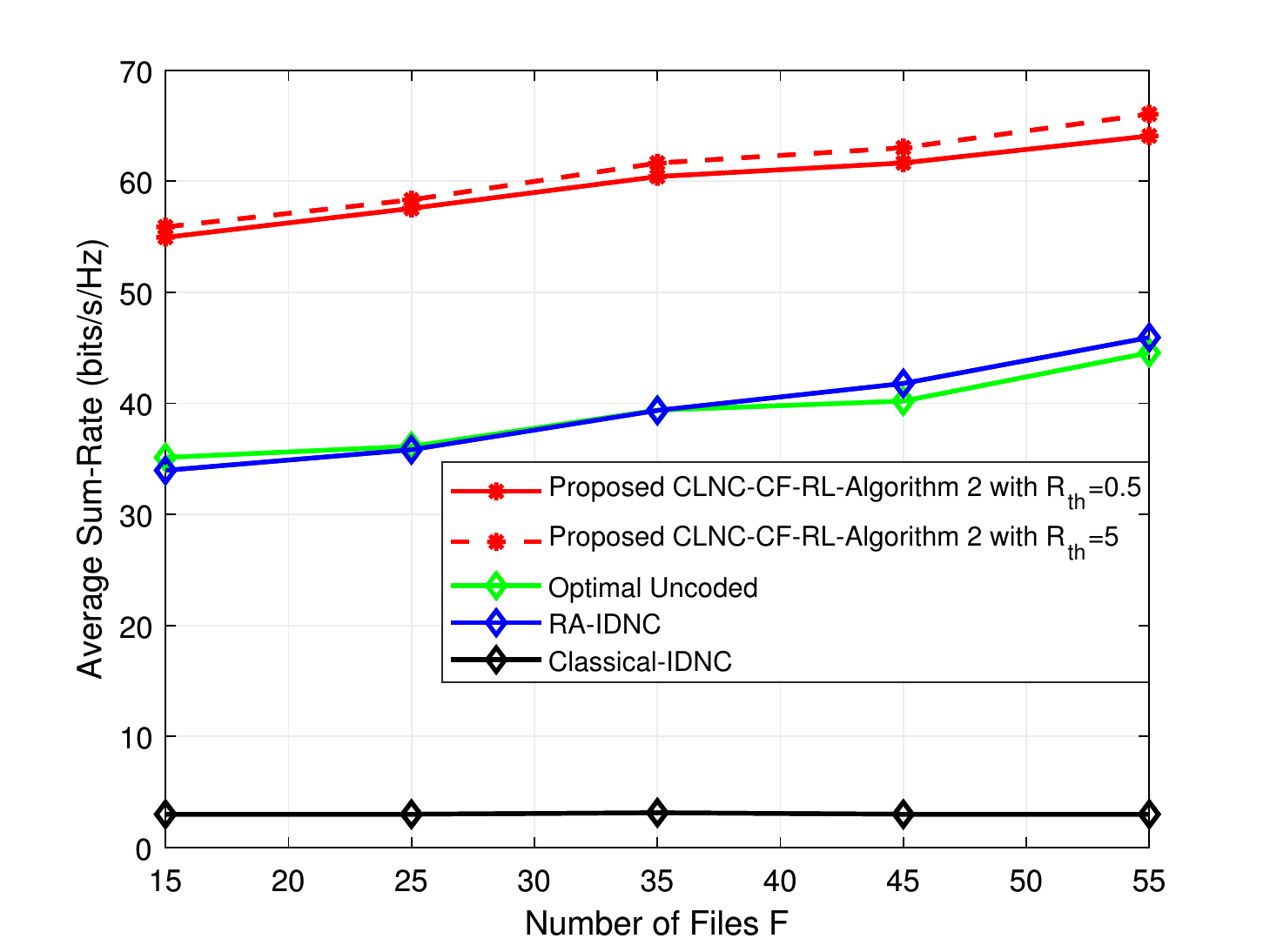} 
		\caption{Sum-rate vs.  number of files $F$ with an optimized caching policy using MARL for $40$ users.}
		\label{fig8}
	\end{minipage}\hfill
\end{figure}

\textit{2) Simulation of Algorithm \ref{Algorithm1}:} In Figs. \ref{fig7} and \ref{fig8}, we demonstrate the effectiveness of the proposed MARL algorithm that efficiently optimizes the caching policy on improving sum-rate vs. number of users and files.  First, we can see that the sum-rate performances of all schemes in Figs. \ref{fig7} and \ref{fig8} are better than that in Figs. \ref{fig3} and \ref{fig4}. Recall, in Figs. \ref{fig3} and \ref{fig4} the random caching policy are used. This shows the effectiveness of the proposed MARL algorithm that works as a key factor to make proper caching decisions. Second, when MARL algorithm is jointly developed with CLNC scheme as in our proposed CLNC-CF-RL scheme, the resultant sum-rate performance is superior as compared to all other schemes, which can be observed from Figs. \ref{fig7} and \ref{fig8}. The joint optimization of our proposed CLNC-CF-RL scheme has several potentials: i) it considers the caching optimization objective using MARL algorithm, 2) it considers the cluster formation behavior of F-APs and CE-D2D users by learning from the history environment, which is the key to make proper caching decisions, and 3) it utilizes the CLNC optimization for NC user scheduling and power levels control. 

\begin{figure}[t!]
	\centering
	\begin{minipage}{0.494\textwidth}
		\centering
		\includegraphics[width=0.85\textwidth]{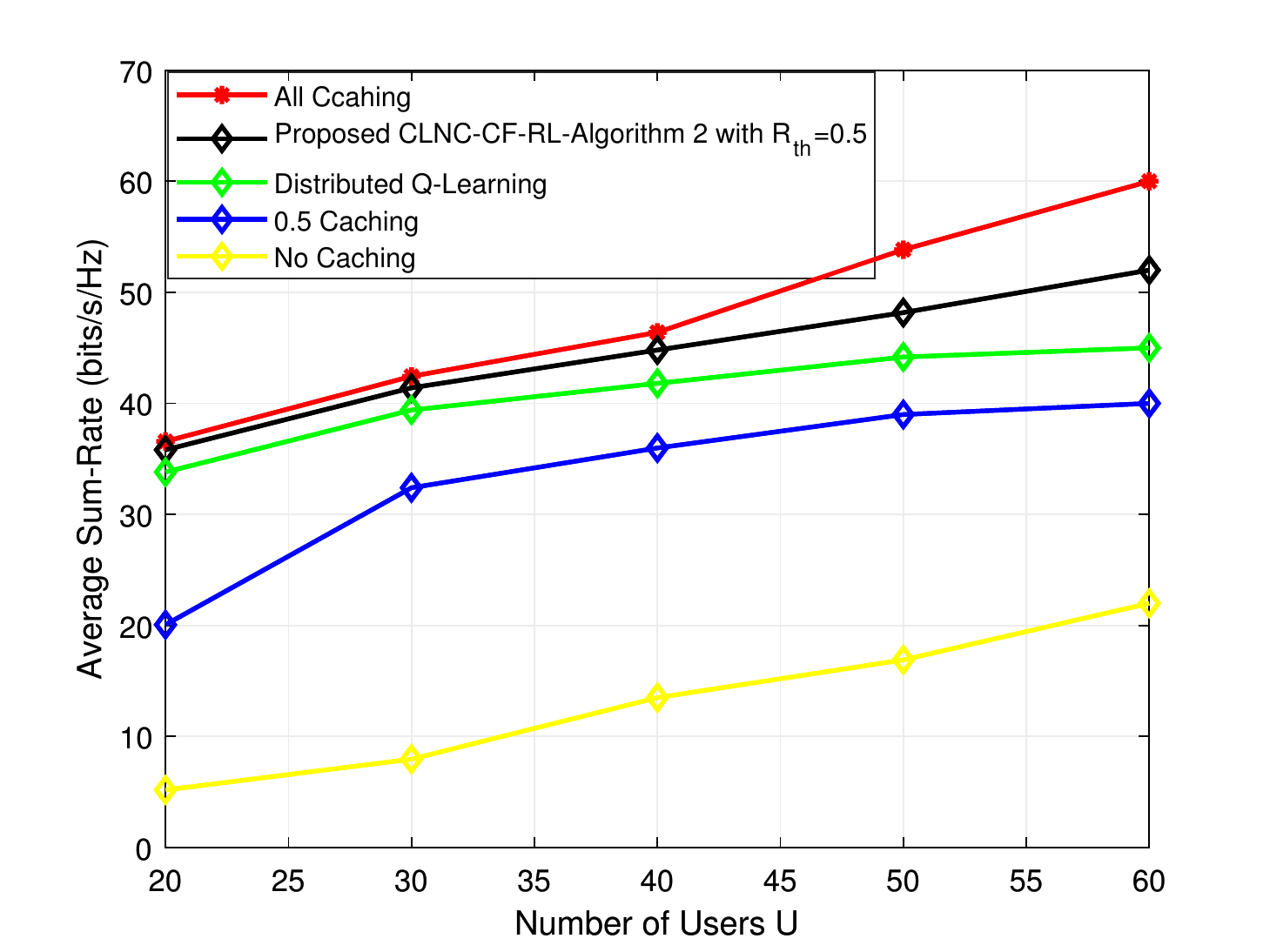} 
		\caption{Sum-rate vs.  number of users $U$ with an optimized caching policy using MARL for $30$ files.}
		\label{fig9}
	\end{minipage}\hfill
	\begin{minipage}{0.494\textwidth}
		\centering
		\includegraphics[width=0.85\textwidth]{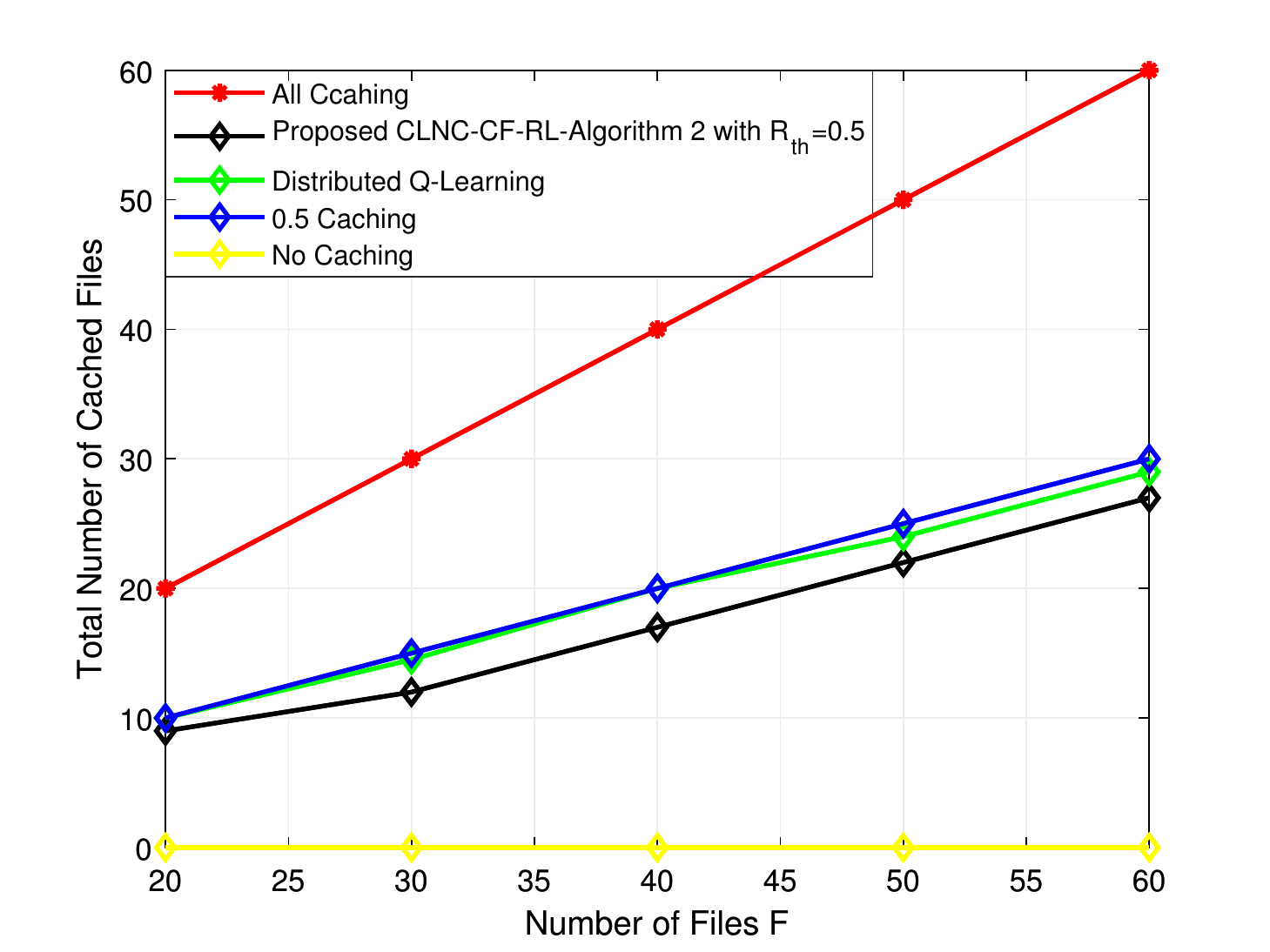} 
		\caption{Total number of cached files vs.  number of files $F$ with an optimized caching policy using MARL for $30$ users.}
		\label{fig10}
	\end{minipage}\hfill
\end{figure}

In Fig. \ref{fig9}, we present the effectiveness of our proposed MARL algorithm on improving sum-rate vs. number of users under different caching schemes.  According to Fig. \ref{fig9}, our proposed scheme outperforms
the no caching, $0.5$ caching, and distributed Q-learning schemes. This is due to the fact that our proposed scheme optimizes the agents' actions and the cluster formation
behavior of F-APs and CE-D2D users by learning from the feedback signal of the
history environment, which helps making proper caching decisions. Particularly, the all caching scheme caches all the requested popular files at F-APs and CE-D2D users, and accordingly, it fully exploits the cooperation between F-APs and CE-D2D users that significantly improves system sum-rate. However, each F-AP and CE-D2D user in this scheme needs to cache all the files, which is impractical. For no caching scheme, the system sum-rate is the worst since there are less F-AP and CE-D2D users' cooperation chances under a limited fronthaul capacity. Compared to the distributed Q-learning scheme, our proposed scheme improves the sum-rate by 10.3\%, which again confirms its superiority. Finally, in Fig. \ref{fig10}, we show that our proposed scheme is also efficient in terms of the total number of cached files as compared to all other schemes.

\section{Conclusion} \label{C}

In this paper, a joint cache and radio resource allocation optimization problem has been evaluated and studied for F-RAN system with D2D communications, which includes a long-term caching optimization problem and an instantaneous cluster formation problem. The considered joint problem has been modeled as a Stackelberg game between the multi-agent leaders, and the clustering behavior of F-APs and CE-D2D users is captured by a CFG. To achieve Stackelberg equilibrium, a distributed CLNC clustering formation algorithm is first developed for F-APs to reach a stable solution under any fixed caching strategy. Then, to tackle the challenges incurred by no closed form and binary caching decision variables, an innovative MARL algorithm is developed to achieve a local optimal caching strategy. Simulation results showed that the proposed joint CLNC-CF-RL framework can effectively improve the sum-rate by around 30\%, 60\%, and 150\%, respectively, compared to: 1) an optimal	uncoded algorithm, 2) a standard RA-IDNC algorithm, and 3) a benchmark classical IDNC with network-layer optimization.

	\newpage
	\begin{IEEEbiography}
		[{\includegraphics[width=1in,height=1.25in,clip,keepaspectratio, ]{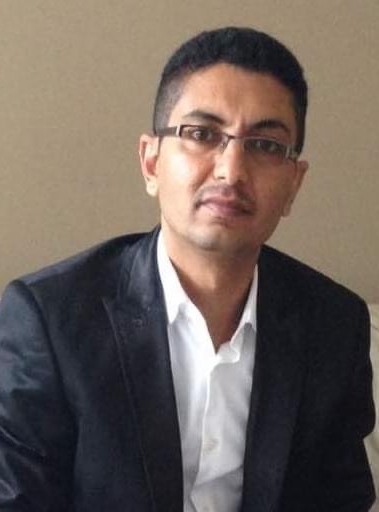}}]{Mohammed S. Al-Abiad} received the B.Sc. degree in computer and communications engineering from Taiz University, Taiz, Yemen, in 2010, the M.Sc. degree in electrical engineering from King Fahd University of Petroleum and Minerals, Dhahran, Saudi Arabia, in 2017, and the Ph.D. degree in electrical engineering from the University of British Columbia, Kelowna, BC, Canada, in 2020.  He is currently a Postdoctoral Research Fellow with the School of 	Engineering at the University of British Columbia, Canada. His research interests include  cross-layer network coding, optimization and
		resource allocation in wireless communication
		networks,  machine learning, and game theory. He is a student member
		of the IEEE.
	\end{IEEEbiography}
	
	\vspace{-1.5cm}	
		\begin{IEEEbiography}
		[{\includegraphics[width=1in,height=1.25in,clip,keepaspectratio, ]{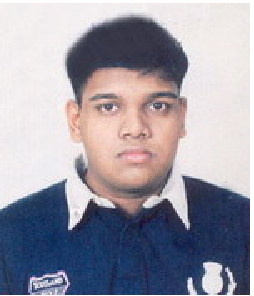}}]{Md. Zoheb Hassan}  received
		the Ph.D. degree from the University of British
		Columbia, Vancouver, BC, Canada, in 2019. He is
		a Research Fellows with the $\acute{\text{E}}$cole de technologie sup$\acute{\text{e}}$rieure (ETS), University of Quebec, Canada. His research interests include
		wireless optical communications, optimization and
		resource allocation in wireless communication
		networks, and digital communications over fading
		channels. He was the recipient of Four-Year Doctoral
		Fellowship of the University of British Columbia in
		2014. He serves/served as a Member of the Technical Program Committee of
		IEEE IWCMC 2018, IEEE ICC 2019, and IEEE ICC 2020.
	\end{IEEEbiography}

	\vspace{-1.5cm}	
	\begin{IEEEbiography}
		[{\includegraphics[width=1in,height=1.25in,clip,keepaspectratio]{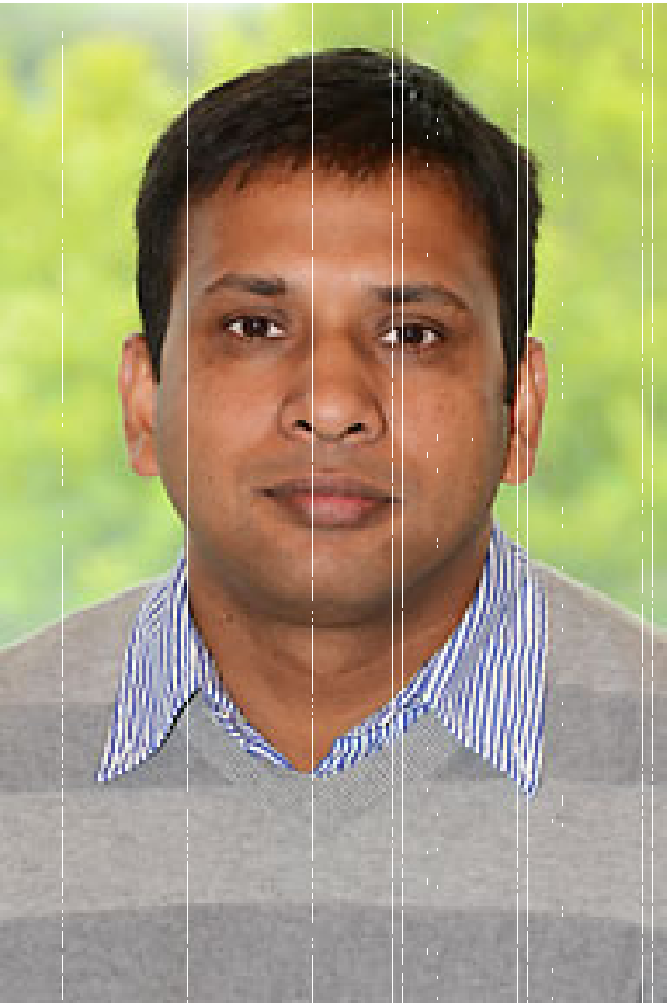}}]{Md. Jahangir Hossain}(S’04, M’08, SM’18) received the B.Sc. degree in electrical and	electronics engineering from the Bangladesh	University of Engineering and Technology (BUET),	Dhaka, Bangladesh, the M.A.Sc. degree from	the University of Victoria, Victoria, BC, Canada,
		and the Ph.D. degree from The University of
		British Columbia (UBC), Vancouver, BC. He was
		a Lecturer with BUET. He was a Research Fellow
		with McGill University, Montreal, QC, Canada,
		the National Institute of Scientific Research,
		Quebec, QC, and the Institute for Telecommunications Research, University
		of South Australia, Mawson Lakes, Australia. His industrial experience 	includes a Senior Systems Engineer position with Redline Communications,
		Markham, ON, Canada, and a Research Intern position with Communication 	Technology Lab, Intel, Inc., Hillsboro, OR, USA. He is currently an Associate Professor with the School of Engineering, UBC Okanagan campus, Kelowna,	BC. His research interests include designing spectrally and power-efficient	modulation schemes, applications of machine learning for communications,	quality-of-service issues and resource allocation in wireless networks, and
		optical wireless communications. He regularly serves as a member of the Technical Program Committee of the IEEE International Conference
		on Communications (ICC) and the IEEE Global Telecommunications	Conference (Globecom). He has been serving as an Associate Editor for
		IEEE COMMUNICATIONS SURVEYS AND TUTORIALS and an Editor for 	IEEE TRANSACTIONS ON COMMUNICATIONS. He previously served as an
		Editor for IEEE TRANSACTIONS ON WIRELESS COMMUNICATIONS.
	\end{IEEEbiography}

\end{document}